\documentclass[12pt]{article}
\pdfoutput=1
\usepackage{natbib}
 \bibpunct{(}{)}{;}{a}{,}{,}
\usepackage{graphicx}
\usepackage{amsmath,amssymb,amsthm}
\usepackage{bm}
\usepackage{rotating}
\usepackage[lined,ruled]{algorithm2e}
\usepackage{multirow}
\usepackage{rotating}
\usepackage{multicol}
\usepackage{titlesec}
\usepackage{latexsym}
\usepackage[pdftex,colorlinks=true,linkcolor=blue,citecolor=blue,urlcolor=blue,bookmarks=false,pdfpagemode=None]{hyperref}
\usepackage{url}
\usepackage{verbatim}
\usepackage{fancyhdr}
\usepackage{setspace}
\usepackage{paralist}
\usepackage{rotating}
\usepackage{lineno}
\usepackage[top=1in, bottom=1in, left=1in, right=1in]{geometry}
\usepackage{lscape}
\usepackage{dcolumn}
\usepackage{color}

\newcolumntype{.}{D{.}{.}{-1}}

\def\abovestrut#1{\rule[0in]{0in}{#1}\ignorespaces}
\def\belowstrut#1{\rule[-#1]{0in}{#1}\ignorespaces}
\def\abovespace{\abovestrut{0.20in}}

\def\belowspace{\belowstrut{0.10in}}

\newtheorem{theorem}{Theorem}
\newtheorem{corollary}[theorem]{Corollary}

\usepackage{caption}

\usepackage{subfigure}
\providecommand{\U}[1]{\protect\rule{.1in}{.1in}}
\renewcommand{\vec}[1]{\boldsymbol{\mathbf{#1}}}

\ifdefined\revision
  \newcommand{\revs}[1]{\textcolor{red}{#1}}
\else
  \newcommand{\revs}[1]{#1}
\fi

\ifdefined\revision

\else

\fi

\ifdefined\revision
  \newcommand{\edits}[1]{\textcolor{red}{\st{#1}}}
\else
  \newcommand{\edits}[1]{}
\fi

\begin{document}
\pagestyle{empty}

\title{
Non-standard conditionally specified models for non-ignorable missing data 
\protect\thanks{Alexander M. Franks is a Moore-Sloan Data Science Fellow at the University of Washington and a graduate of the Department of Statistics at Harvard University (\href{mailto:afranks@post.harvard.edu}{afranks@post.harvard.edu})
Edoardo M.~Airoldi is an Associate Professor of Statistics at Harvard University (\href{mailto:airoldi@fas.harvard.edu}{airoldi@fas.harvard.edu}). Donald B. Rubin is the John L. Loeb Professor of Statistics at Harvard University (\href{mailto:dbrubin@fas.harvard.edu}{dbrubin@fas.harvard.edu})
This work was partially supported 
 by the National Science Foundation under grants 
  CAREER IIS-1149662 and IIS-1409177,
 and by the Office of Naval Research under grant 
  YIP N00014-14-1-0485. 
 Edoardo M.~Airoldi is an Alfred Sloan Research Fellow, and a Shutzer Fellow at the Radcliffe Institute for Advanced Studies.
The authors are grateful to Dr.\ Shahab Jolani (Department of Methodology and Statistics, Faculty of Social Sciences,
Utrecht University) and Dr.\ Stef Van Buuren (Netherlands Organisation for Applied Scientific Research TNO) for sharing preliminary work and analyses that contributed to the framing of this paper.}}

\author{
 Alexander M. Franks, Edoardo M. Airoldi, Donald B. Rubin\\
 Department of Statistics\\
 Harvard University, Cambridge, MA 02138, USA}
\date{}

\maketitle
\thispagestyle{empty}

\newpage
\begin{abstract}
Data analyses typically rely upon assumptions about missingness mechanisms that lead to observed versus missing data. When the data are missing not at random, direct assumptions about the missingness mechanism, and indirect assumptions about the distributions of observed and missing data, are typically untestable. We explore an approach, where the joint distribution of observed data and missing data is specified through non-standard conditional distributions. In this formulation, which traces back to a factorization of the joint distribution, apparently proposed by J.W. Tukey, the modeling assumptions about the conditional factors are either testable or are designed to allow the incorporation of substantive knowledge about the problem at hand, thereby offering a possibly realistic portrayal of the  data, both missing and observed. We apply Tukey's conditional representation to exponential family models, and we propose a computationally tractable inferential strategy for this class of models. We illustrate the utility of this approach using high-throughput biological data with missing data that are not missing at random.

\vfill
\end{abstract}

\newpage
\singlespacing
\small
\tableofcontents
\normalsize
\doublespacing

\newpage
\pagestyle{plain}
\setcounter{page}{1}


\section{Introduction}

Missing data are ubiquitous in  applications of statistics, including survey sampling, healthcare, public policy and bioinformatics.
The effort needed for modeling and analyzing observations in these situations crucially depends on the mechanism that induces the missing data as well as on the mode of inference \citep{littlebook}.

Here, we work within the inferential framework outlined in \citet{rubi:2004}.
The inferential target of interest is generally a function of  observed and missing data.
One key concept is that of  a ``missing at random'' missing data mechanism, for which the conditional probability distribution of the missingness indicators is  a function only of the observed data, and the related concept of ignorable missing data \citep{Rubin1976,rubin1978aa,Mealli:2015aa}. 
For Bayesian inference, ignorability implies that the posterior distribution of the target is conditionally independent
of the observation indicators, given the observed data, and so 
there is no need to specify a model for observation indicators to achieve valid Bayesian or likelihood-based inference. When the posterior distribution of the
inferential target depends on these indicators, valid Bayesian
 inference requires specifying a joint model for the data and
the  indicators.

With missing data, there are two basic
approaches to specify the the joint distribution of the complete data and missing
data indicators given parameters, and the implied likelihood, as a product of marginal and conditional
distributions.
The first basic approach \citep{rubin1974aa} is to posit a standard model for the complete data, and
then specify an explicit model that selects observed data from the complete data, called the missingness mechanism \citep{Rubin1976}. 
The second basic approach is to specify separate distributions for the observed data and the missing data, thus eschewing explicit assumptions about the missingness mechanism \citep{Rubin1977,little1993}.
 The fundamental challenge with these two basic approaches is that
 some assumptions about the missingness mechanism, whether explicit or implicit, are typically not testable from the observed data. 
 As a result, some literature on inference in the presence of missing data  centers on assessing sensitivity to different model specifications \citep{Rubin1977,Scharfstein1999,andrea2001,littlebook}.  

\subsection{Contributions}

Here, we develop an alternative approach to modeling data possibly missing not at random, evidentially originally proposed by Tukey and discussed by Hartigan and Rubin, reported by \citet{holland-notes}.
The key insight is to represent the joint distribution of the complete data and  missing-data indicators as proportional to factors that involve only  observed values and the missingness mechanism.  Assumptions about these factors are either testable, or typically allow the incorporation of substantive knowledge about the problem at hand, thereby offering a clear path to eliciting a realistic portrayal of the sources of variation in the data.

In Section \ref{sec:mnar}, we review the two basic  model specifications for  missing data, before formally introducing Tukey's representation.  In Section \ref{sec:bayesian}, we discuss technical issues involved when using Tukey's representation, and provide a full characterization for exponential-family models. In Section \ref{sec:results}, we then illustrate the use of these models on simulated data when the model is correctly specified,  when the model is incorrectly specified, and on an application in  biology. In Section \ref{sec:disc}, we offer  theoretical insights and use them to state formal results.

\section{Basic models for  missing data}

\label{sec:mnar}

Discussion of  models using basic factorizations for missing data can be found in a variety of places, including \citet{glynn} and \citet{handbook}.

Throughout, let $Y = (Y_1, Y_2, ..., Y_N)^\intercal$ represent the
complete data and $R=(R_1,R_2,...,R_N)^\intercal$ represent the response
indicators for $Y$, which are ``missing'' when $R_i$=0 and
``observed'' when $R_i=1$.  The joint distribution, assuming $(Y_i,R_i)$ are i.i.d., is then
$$P(Y,R \mid \theta) = \prod_i f(Y_i,R_i\mid\theta),$$ 
where $\theta$ is a  parameter vector. 
For simplicity,  we focus on the the case without covariates. 

\subsection{The selection factorization}
\label{sec:selfact}

The selection approach \citep{rubin1974aa} factors the joint distribution of $(Y,R)$ as  
\begin{equation}
\label{eq:selfac}
P(Y,R \mid \theta) = \prod_i^N f(Y_i|\theta_Y)f(R_i|Y_i,\theta_{R|Y}),
\end{equation}
using the  distribution of the complete data, $P(Y|\theta_Y)$, and the missingness mechanism, or selection function, $P(R|Y,\theta_{R|Y})$, which controls
which data are actually observed, where the parameters $(\theta_Y, \theta_{R|Y})$,  are functions of $\theta$.
Models for $f(Y_i|\theta_Y)$ include the normal, with $\theta_Y$ the mean and variance of the normal, or the Bernoulli with $\theta_Y$ the probability of success of the Bernoulli.  Typical models for $f(R_i|Y_i,\theta_{R|Y})$ include the logistic and probit models \citep[e.g., see][]{Gelman2004}.  

\subsection{The pattern-mixture factorization}
\label{sec:patfact}

The pattern-mixture approach \citep{Rubin1977,little1993,littlebook} is the  alternative basic factorization; the complete data distribution is specified as a mixture of observed
and missing data components,
\begin{eqnarray}
\label{eq:patmix}
P(Y,R | \theta) &=& \prod_{i=1}^N f(Y_i|R_i, \theta_{Y|R})f(R_i|\theta_R) \nonumber \\
&=& \prod_{i=1}^N\prod_{r=0}^1 \left[f(Y_i|R_i=r, \theta_{Y|R=r})f(R_i=r|\theta_R)\right]^{I(R_i=r)},
\end{eqnarray}
where $(\theta_{Y|R},\theta_R)$ are functions of $\theta$, leading to a mixture likelihood, which for a single  observation is
$$ P(Y_i \mid \theta) = \sum_{r=0}^1 f(Y_i|R_i=r, \theta_{Y|R=r})f(R_i=r|\theta_R).$$
The  model for $f(R_i|\theta_R)$ is a Bernoulli distribution with parameter $\theta_R$. The model for $f(Y_i|R_i=1)$ is typically chosen to fit the observed data well, whereas the model for $f(Y_i|R_i=0)$ is commonly chosen to be a location shift or scale change of $f(Y_i|R_i=1)$ \citep[e.g., see][]{Gelman2004,littlebook}.

\subsection{Tukey's representation}
\label{sec:tukey}

John W.\ Tukey, in a discussion of  \citet{glynn},
suggested an alternative factorization of the joint distribution for
$(Y,R)$ in terms of conditional distributions \citep[recorded in][]{holland-notes}, which he refers to as the {\em simplified
  selection model}, with parameters $\theta_{Y|R=1}$ and $\theta_{R|Y}$,
\begin{equation}
\label{tukeysFactorization}
P(Y,R \mid \theta ) \propto \prod_{i=1}^N f(Y_i\mid R_i=1, \theta_{Y|R=1}) \cdot \frac{f(R_i\mid Y_i, \theta_{R|Y})}{f(R_i=1\mid Y_i,\theta_{R|Y})},
\end{equation}
with normalizing constant $\prod_{i=1}^N f(R_i=1\mid\theta_{R|Y},\theta_{Y|R=1})$ ensuring integrability. 
\revs{
As Holland notes, a main advantage of this factorization is
that it only involves the observed data density, $f(Y_i \mid R_i=1, \theta_{Y|R=1})$, which can be
estimated directly, and the missingness mechanism, $f(R_i\mid Y_i, \theta_{R|Y})$, which can be easy to elicit in the context of a specific application.}

Tukey's representation can be obtained through a simple application of Brook's
lemma, which equates the ratio of joint distributions to the ratio of their full conditionals \citep{brook64,besag74}.  
Although this lemma is most commonly referenced in the theory of
spatial autoregressive models \citep{cressie}, its connection to
Tukey's representation is relevant because it immediately reveals some theoretical insights.
Importantly, Brook's Lemma is only applicable
when the so-called \emph{positivity condition} is satisfied \citep{hammersley1971aa}, which for Tukey's representation  means that
\begin{align*}
&\text{If: } P(R_i=r|\theta)>0 \text{ and } P(Y_i=y|\theta)>0 \\
&\text{Then: } P(R_i=r,Y_i=y|\theta) > 0
\end{align*}
for all  pairs of values $(r,y)$.
This condition is not trivially satisfied in  missing data problems.  For instance, Tukey's representation {cannot} be applied to models where $P(R_i=1|Y_i<c,\theta_{R|Y}) = 0$, deterministically, for some cutoff $c$, as when the complete data model is normal and the observed data model is truncated normal.  Consequently, here we focus on problems where $P(R_i=1|Y_i,\theta_{R|Y}) > 0$, that is, where the support of the missing data is contained in the support of the observed data.

In addition, the  conditional distributions specified in Equation \ref{tukeysFactorization}
must imply an integrable joint density \citep{besag74}.  With
Tukey's representation, the \emph{integrability condition} constrains the rate
at which the tails of the distribution for observed data decrease relative to the rate
at which the odds of a missing value increase. This condition is illustrated in Section \ref{sec:sim}, and discussed in Section \ref{sec:integrability}.







\section{Modeling and inference using Tukey's representation}
\label{sec:bayesian}

Let $P(Y_i = y_i \mid R_i=1,\theta_{Y_i|R_i=1})$ be denoted $f^{\rm obs}(y_i\mid\theta_{Y|R})$, for simplicity.
Using Equation \ref{tukeysFactorization}, we can write the density for $(y_i,r_i)$ as
\begin{eqnarray}
\label{lik:comp}
f(y_i,r_i \mid \theta_{R|Y},\theta_{Y|R}) 
&\propto& \begin{cases} f^{\rm obs}(y_i\mid\theta_{Y|R}) &\mbox{if } r_i=1 \nonumber \\ 
\frac{f(r_i=0\mid y_i, \theta_{R|Y})}{f(r_i=1\mid y_i, \theta_{R|Y})} f^{\rm obs}(y_i\mid\theta_{Y|R}) & \mbox{if } r_i=0\\
\end{cases} \\
&=& f(r_i=1\mid y_i, \theta_{R|Y})^{r_i-1}f(r_i=0\mid y_i, \theta_{R|Y})^{1-r_i}f^{\rm obs}(y_i\mid\theta_{Y|R})
\end{eqnarray}
with normalizing constant 
%
\begin{eqnarray}
\label{eqn:normconst}
Q (\theta_{Y|R},\theta_{R|Y}) 
 &=& \left(1+\int \frac{f(r_i=0\mid y_i, \theta_{R|Y})}{f(r_i=1\mid y_i, \theta_{R|Y})} f^{\rm obs}(y_i \mid \theta_{Y|R}) ~dy_i\right)^{-1},
\end{eqnarray}
which ensures the integral of Equation \ref{lik:comp} over  random variables $R$ and $Y$ times $Q$ is unity.

The normalizing constant $Q$  is generally difficult to compute. As  a consequence, the missing data density   cannot easily be expressed.    Below, we introduce a class of models for which computation of  the normalizing constant is tractable and which also implies simple  distributional forms for the missing data and complete data  densities.

\subsection{Exponential family models}
\label{sec:ef}

Assume that the observed data distribution belongs to an exponential family and that 
the logit of the missingness mechanism is linear in the sufficient statistics of that family, which is related to the
exponential tilt pattern-mixture models introduced by
\citet{Birmingham2003}.  
Formally, let $f^{\rm obs}(y_i \mid \theta_{Y|R=1})$ be an exponential family distribution with natural parameter $\theta_{Y|R=1}$.  Then, 
\begin{equation}
\label{eq:exp-log-obs}
f^{\rm obs}(y_i \mid \theta_{Y|R=1}) = h(y_i)g(\theta_{Y|R=1})e^{T(y_i) ' \theta_{Y|R=1}},
\end{equation}
where $g(\theta_{Y|R=1})$ is the normalizing constant and $T(y_i)$ is the natural sufficient statistic, possibly multivariate.
The missingness mechanism
\begin{equation}
\label{eq:exp-log-sel}
f(r_i = 1 \mid y_i,\theta_{R|Y}) = \hbox{logit }(T(y_i)'\theta_{R|Y}) = \frac{1}{1+e^{- \theta_{R|Y}}}
\end{equation}
implies that
\begin{equation}
\nonumber \frac{f(r_i = 0 \mid y_i,\theta_{R|Y})}{f(r_i = 1 \mid y_i,\theta_{R|Y})} = e^{- T(y_i)\theta_{R|Y}}.
\end{equation}
Then the normalizing constant $Q$ in Equation \ref{eqn:normconst} can be written as a simple function of the normalizing constant $g(\cdot)$ in the exponential family formulation of $f^{\rm obs}$,
\begin{eqnarray}
\label{eqn:exfam} 
  Q(\theta_{Y|R=1},\theta_{R|Y}) 
  &=& \frac{g({\theta_{Y|R=1}}+{\theta_{R|Y}})}{g({\theta_{Y|R=1}}+{\theta_{R|Y}})+g(\theta_{Y|R=1})}.
\end{eqnarray}

For the class of exponential-logistic models defined by Equations \ref{eq:exp-log-obs}--\ref{eq:exp-log-sel}, the missing data distribution, as specified in Equation \ref{miss_dist_ef}, is from the same exponential family as the
observed data with natural parameter\footnote{In the exponential-logistic models we consider in this section, the two parameter vectors $(\theta_{Y|R=1},\theta_{R|Y})$ always have the same dimensionality. }
$\theta_{Y|R=0} = \theta_{Y|R=1} + \theta_{R|Y}$,
\begin{equation}
\label{miss_dist_ef}
f^{\rm mis}(y \mid \theta_{Y|R=1},\theta_{R|Y}) = h(y) g(\theta_{Y|R=0}) e^{{T(y)}'\theta_{Y|R=0}}.
\end{equation}
Thus, missing data imputation with this class of models is straightforward.  

In Tukey's representation,
 the main source of uncertainty about any inferential target is due to the missingness mechanism, and the prior distribution on $\theta_{R|Y}$.

\subsection{Estimation and inference}
\label{sec:est:inf}

The primary estimands of interest are typically functions of the
parameters specifying the complete-data distribution, $(\theta_{Y|R=1},\theta_{R|Y})$.  
Because the observed and missing data densities for exponential-logistic models are exponential families, see Equations \ref{eq:exp-log-obs} and \ref{miss_dist_ef},  the complete-data distribution is a mixture of exponential
families.  Further, analytic expressions for the normalizing constant
$Q$ and the likelihood are available, and thus standard Markov Chain
Monte Carlo methods are applicable \citep{RoberCase}. 

We take a simple approach to inference via Markov Chain Monte Carlo, which is computationally less
demanding than with samplers that explicitly characterize the geometry of the solution space, as discussed in Section \ref{sec:disc}.  
Consider a simple Normal-logistic model as an illustration, with  $\theta_{R|Y} = (\beta_0,\beta_1) = \beta
\in R^2$ with $\beta_0$ corresponding to the intercept and $\beta_1$
the rate at which the odds of selection change in $T(y)=y$:
\begin{eqnarray}
 f(r_i = 1 \mid y_i,\beta) &=& \hbox{logit }(\beta_0+\beta_1 y_i) 
                            = \bigm(1+\exp\{- \beta_0-\beta_1 y_i\}\bigm)^{-1} \nonumber \\
 f^{\rm obs}(y_i)          &=& \hbox{Normal }(0,1). \nonumber
\end{eqnarray}
Rather than specifying a
prior distribution on  $(\beta_0, \beta_1)$, we specify a prior distribution on $Q$ and 
$\beta_1$, but not on $\beta_0$. We then invert Equation \ref{eqn:exfam} and solve for $\beta_0$ to get,
\begin{equation}
\label{eqn:invert}
\beta_0(Q,\eta,\beta_1) =
\text{log}\left(\frac{g(\eta+\beta_1)(1-Q)}{g(\eta)}\right).
\end{equation}
In this Normal-logistic example, 
the standard Normal has natural parameters $(\eta_1,\eta_2)=(0,-1/2)$, and  $g(\eta) =
-\frac{\eta_1^2}{4\eta_2} - \frac12\ln(-2\eta_2)$.  Thus, Equation \ref{eqn:exfam} becomes,
\[
 Q(\beta_0,\beta_1) = \frac{\beta_1^2}{\beta_1^2-2e^{\beta_0}},
\]
and solving Equation \ref{eqn:invert} we get
\[
\beta_0 = \text{log}\left(\frac{\beta_1^2(1-Q)}{2}\right).
\]
Next, we use this inferential strategy to demonstrate the utility of Tukey's representation.

\section{Numerical examples and an application in biology}
\label{sec:results}

In Section \ref{sec:sim}, we explore selection, pattern-mixture, and Tukey representations for missing data using simulation studies, where $f^{\rm obs}(y_i\mid r_i=1,\theta_{Y|R=1})$ is a mixture of exponential family distributions.  
In this example, we discuss how both the selection and the pattern mixture factorizations require difficult modeling choices in practice, and illustrate how Tukey's representation can be employed with relative ease.  
In Section \ref{sec:robustness} we explore the robustness of logistic-exponential family missing-data models when the true generating process cannot be expressed as a finite mixture of exponential families.
In Section \ref{sec:app} we demonstrate the utility of Tukey's factorization in an application to biological data \citep{franks2014}.


\subsection{Illustration with semicontinuous data}
\label{sec:sim}

Let us posit that the complete data observations are drawn from a mixture of discrete and continuous distributions, which we term a semicontinuous distribution.  Specifically, we assume that the continuous component is a mixture of normals and that the missing data mechanism is logistic in the data.  Here, we specify the full data generating process using Tukey's representation.  First, we assume the observed data density has form
\[
 f^{\rm obs}(y_i \mid r_i=1, \theta_{Y|R=1}) = \alpha\left(\sum_{k=1}^K w_k \cdot \hbox{Normal }(y_i; \mu_k,\sigma_k^2)\right)+(1-\alpha)\left(\sum_{m=1}^M \, p_m \cdot \delta_{\{\gamma_m\}}(y_i)\right)
 \]
 where $\theta_{Y|R=1} = (\alpha, w_{1:k,}, \mu_{1:k}, \sigma^2_{1:k}, p_{1:m}, \gamma_{1:m})$, and $\delta_{\{\gamma\}}(y)$ is the Dirac delta function that is one when $y=\gamma$, and zero elsewhere.  The parameter $\alpha$ captures the fraction of observations attributable to the continuous component, and the parameters $p_{1:M}$ specify the probabilities for the point masses at $M$ locations $\gamma_{1:M}$.  Recalling the notation introduced in Section \ref{sec:ef}, for the $k$-th Normal mixture component, the natural parameter is $\theta_{Y|R=1} = (\mu_k,\sigma^2_k)$ and the corresponding sufficient statistics is $T(y_i) = (y_i,y_i^2)$. For the discrete distribution, $\theta_{Y|R=1} = p_{1:M}$ and $T(y_i) = 1$ if $y_i$ equals one of the $M$ discrete locations, zero otherwise.

The missingness mechanism specifies that the logit of the selection probabilities are
quadratic in $y_i$, but importantly, linear as a function of the sufficient statistics of the normal components:
\begin{eqnarray}
\label{sim:mech}
f(r_i=1 \mid y_i,\theta_{R|Y})  
   &=& \hbox{logit}^{-1}(-(\beta_0 + (y_i-\beta_1)^2\beta_2)) \\
   &=& \bigm(1+\exp\{\beta_0 + y_i^2\beta_2 - 2 \beta_1 \beta_2 y_i + \beta_1^2 \beta_2\}\bigm)^{-1} \nonumber 
\end{eqnarray}
where $\theta_{R|Y} = (\beta_0,\beta_1,\beta_2)$.  Here, $\beta_1$ corresponds to the value at which an observation is most likely to occur, whereas $\beta_2$ controls how quickly the observation probability decays with the distance of $y_i$ to $\beta_1$.  Non-monotonicity in the probabilities of selection occurs in practice when extreme values, both small and large, are less likely to be observed.

Using Equation \ref{miss_dist_ef}, we can see that  the missing
data distribution for the continuous component is also a  mixture of Normals.  The missing data has density
\begin{eqnarray}
\label{sim:miss}
 f^{\rm mis}(y_i \mid -) 
  &=& \alpha^*\left(\sum_{k=1}^K \, w_k^* \cdot \hbox{Normal }(y_i ; ~\frac{\mu_k - 2 \beta_1 \beta_2 \,\sigma_k^2}{1 - 2\beta_2 \,\sigma_k^2}, ~\frac{\sigma_k^2}{1 - 2 \beta_2 \, \sigma_k^2})\right) ~+ \nonumber \\
  &+& (1-\alpha^*)\left(\sum_{m=1}^M \,p^*_k \cdot \delta_{\{\gamma_j\}} (y_i)\right).
\end{eqnarray}
This density, with specifications for the mixture
weights $\alpha^*$ and $w^*_{1:K}$, is derived in Appendix
\ref{app:sec:sim}.  The observed data and missing data densities specify
the complete data as semi-continuous mixture, from which all relevant
estimands can be computed.

In order to simulate data from this complex mixture, we fix $K=3$ components for the continuous portion of the observed data mixture, with parameters  set as follows,
\begin{align}
\mu_{1:3} & =(-2,0,3)\\
w_{1:3} & =(0.3,0.4,0.3)\\
\sigma_{1:3} & =(1,1,1)~.
\end{align} 
We set $\alpha=0.8$, that is, 80\% of the observed data comes from  the
continuous component. We and fix the discrete component of the observed data to be uniform on $\gamma_{1:9} = \{-4,-3,...,3,4\}$, where $M=9$.  
Finally, for the
missingness mechanism, we assume that $\beta_2 = .06$, $\beta_1=-2$ and
$Q = 0.5$, that is, $50\%$ of the complete data is missing.  We then use
Equation \ref{eqn:invert} to derive that $\beta_0 \approx -0.85$. 
The full derivation is provided in Appendix \ref{app:sec:sim}.

The top panel of Figure
\ref{fig:sim_dens} shows the true missingness mechanism used to
simulate data, as a bold black line.
The bottom panel shows a
histogram of draws from the observed data as well as the true
observed, missing and complete data densities for the continuous
component.  For this simulation study, we will assume that the
relevant estimands are the complete data mean and complete data standard deviation.

\begin{figure}[t!]
     \centering
     \includegraphics[width=0.7\textwidth]{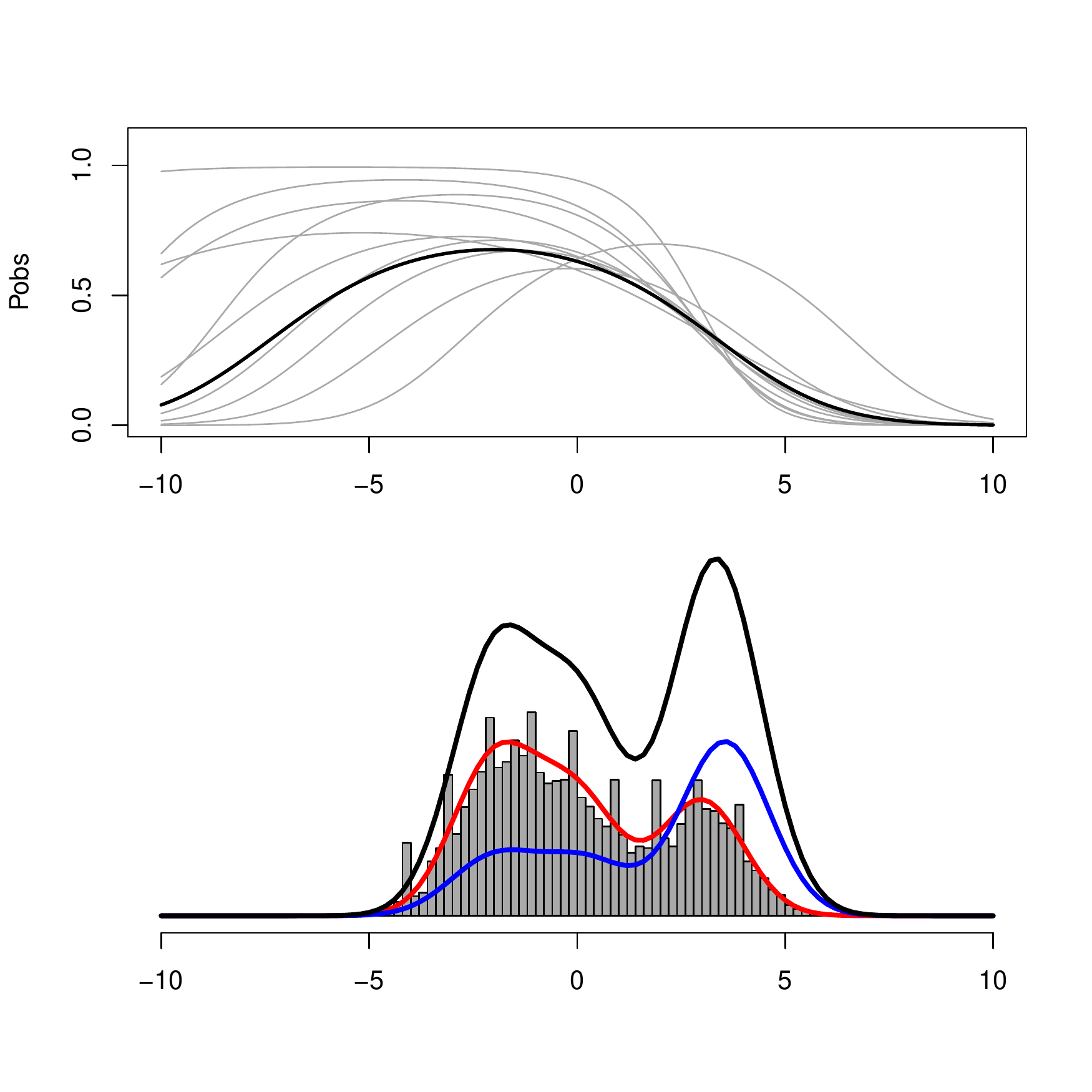}
     \caption{\onehalfspacing Top: the true missingness mechanism (black) and 10
       samples from the prior distribution specified in Section
       \ref{sec:tukeyanalysis} (gray).  Bottom: the observed data
       histogram (gray, $N=10,000$) with observed (red) and missing
       data (blue) densities for the continuous component.  Ten
       percent of the observed data comes from a discrete distribution
       on \{-4,-3,...,0,...,3,4\}.}
\label{fig:sim_dens}
\end{figure}


\subsubsection{Analysis using Tukey's representation}
\label{sec:tukeyanalysis}

Here we use Tukey's representation to estimate the
posterior distribution of the complete data mean and standard
deviation, from observations and observation indicators generated from the
model  above.  In order to do so, we  specify prior distributions for the
parameters of the missingness mechanism, $\beta$, the fraction of
missing data, $Q$, and a model for the observed data, $f^{\rm obs}$. 

We posit the following prior distributions on the parameters of the missingness mechanism,
\begin{eqnarray}
\label{sim:selprior}
\nonumber \beta_1 &\sim& \text{Normal }(-2;2)\\
\nonumber \beta_2 &\sim& 0.08 \cdot \text{Beta }(3;1)\\
Q &\sim& \text{Uniform }(0,1) 
\end{eqnarray}
These prior distribution are centered around the parameters of the true missingness mechanism, $(\beta_1,\beta_2)$, with some variability to reflect a  degree of uncertainty.   Importantly, neither $\beta_1$ nor $\beta_2$ can be estimated from the observed data, and thus the uncertainty in this prior specification will propagate to the posterior, regardless of the amount of observed data.  
The top panel of Figure \ref{fig:sim_dens}  shows a number of draws from the above prior distributions, in gray.  

Next, we posit the following prior distributions for the observed data parameters, $\theta_{Y|R=1}$:
\begin{eqnarray}
\label{sim:selprior}
\nonumber p_i &\sim& \text{Dirichlet}(1, 1, \ldots, 1) \hbox{ for } i=1,\dots,M=9\\
\nonumber \mu_{k} &\sim& \text{Normal}(0, 10) \hbox{ for } k=1,\dots,K=3\\
\nonumber w_{k} & \sim& \text{Dirichlet}(1, 1, 1) \\
 \sigma_{k} & \sim& \text{Uniform}(0, 2)   
\end{eqnarray}
Unlike the prior specification for the parameters of the missingness mechanism, all of the observed data parameters are  estimable.  Thus, the results are only sensitive to the prior specifications of $\theta_{Y|R=1}$ for smaller  observed data sample sizes.  

Note that by the integrability condition stated in Section \ref{sec:ef}, in order for the continuous mixture components of the complete data to all have positive variance, it must hold that $\beta_2 < \underset{k=1 \dots K}{\text{min}} ~\frac{1}{2\sigma_k^2} ~.$  This inequality is due to Equation \ref{sim:mis_moms} in  Appendix \ref{app:sec:sim}.  
To ensure the integrability condition is satisfied, we also bound the
prior distributions on $\sigma_{1:3}$ above by 2, because $0.08 <
\frac{1}{8}$, where $0.08$ is the maximum value of $\beta_2$ under our
prior distribution.

\begin{figure}[t!]
     \centering
     \includegraphics[width=0.95\textwidth]{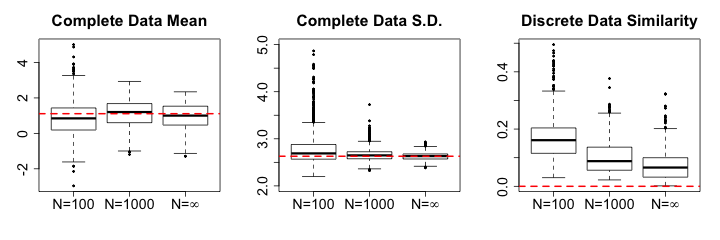}
     \caption{\onehalfspacing Posterior estimates of the complete-data mean (left), complete-data standard deviation (middle) and maximum absolute difference between the true and inferred discrete probability masses (right).  Dashed red line represents the true complete-data value.  As we get more data, estimates of the parameters underlying the observed-data distribution improve, and the posterior concentrates around the true values.  However,  information about the parameters underlying the selection mechanism does not accumulate, thus posterior uncertainty about the complete-data parameters remains.} 
    \label{fig:tukey_sim}
\end{figure}

We present the simulation results for different  sample sizes in
Figure \ref{fig:tukey_sim}.  With a reasonably informative prior distribution for the missingness mechanism, these results show that the analysis with the Tukey's representation accurately recovers the posterior distributions for the relevant estimands.  As the sample size increases, our posterior uncertainty shrinks, but never disappears.  This is because even with perfect knowledge of the observed data density\footnote{\onehalfspacing We simulate this situation  by setting the parameters underlying the observed data distribution to their true values, and we label the corresponding results in Figure \ref{fig:tukey_sim} with $N=\infty$.}, there is no information in the data about the parameters of the missingness mechanism, as mentioned above. Therefore uncertainty about the missingness mechanism propagates to the uncertainty about the complete-data quantities, and is the main source of uncertainty in the posterior when enough data is available. This fact facilitates the sensitivity analysis about the missingness mechanism when using Tukey's representation.   

Above, we illustrated how data analysis using Tukey's representation can be applied in a straightforward manner to a complex data generating process. It
does not require any explicit assumptions about the complete-data
distribution; rather it only requires a reasonable model for the observed data,
and plausible assumptions about the selection mechanism.
In contrast, there are often significant challenges for data analysis when applying the
selection factorization, or pattern mixture factorization, as we now discuss.

\subsubsection{Remarks on the use of the selection factorization}


Inference under selection factorization
models typically require numerical or Monte Carlo integration of the
complete data distribution, and hence is computationally and
analytically more demanding than both pattern mixture models and Tukey's representation.
For instance, often the missing data density does not take on a simple form, and
thus missing data imputation is non-trivial.  Finally, it is not
 obvious that the specified selection factorization model
will imply a reasonable fit to the this observed data.  This is a
fundamental challenge associated with any approach that does not
directly model the observed data \citep{rubi:2004,littlebook,lunagomez2014bayesian,Mealli:2015aa}. The focus on modeling the observed data explicitly, rather than implicitly via assumptions on the data generating process, is one of the strengths of analyses based on the Tukey's representation.

Even if we knew the complete data distribution corresponded to a six
component mixture of Normal with discrete masses at certain values, it
would still be difficult to specify appropriately informed prior
distribution for the relevant parameters. Note that in this example,
the parameters of the missing data density are a function of the
selection parameters (Equation \ref{sim:miss}).  For this reason,
specifying independent prior information for the mixture components of
the complete data density implicitly adds additional prior information
about the parameters of the missingness mechanism.  By parameterizing the
model in terms of complete data parameters $\theta_Y$, we are
combining what we don't have information about ($\theta_{Y|R=0}$) with
what we do have information about ($\theta_{Y|R=1}$)
\citep{holland-notes}. 

\subsubsection{Remarks on the use of the pattern-mixture factorization}

The true data generating process described at the beginning of Section \ref{sec:sim} is specified as a mixture. Thus, at least in principle, the pattern mixture factorization, described in Section \ref{sec:patfact}, provides an appropriate way to model the complete data.  
However, the implied missing data distribution in this example does not correspond to a simple location or scale change of the observed data distribution.  Thus, as is common with pattern mixture models, the missing data distribution for this data cannot be represented with a simple between-components location and scale change \citep[as discussed, e.g., in][]{Daniels2000}.

More generally, when the missing data distribution is complicated, it can be difficult to propose plausible, scientifically justifiable prior distributions for the parameters of the missing data density, $\theta_{Y|R=0}$. 
Perhaps even more problematic is that these parameters often imply smoothness and monotonicity of the selection mechanism, which is not explicitly specified in analyses that use the pattern-mixture factorization.

\subsection{Robustness to Misspecification}
\label{sec:robustness}

In Section \ref{sec:tukeyanalysis}, we performed the analysis using Tukey's representation using a family of models that included the true data generating process.  In practice, finite mixtures of exponential families can only be expected to approximate the observed data density.  Here, we evaluate the robustness of Tukey's approach in such cases. 

We consider  data generated under a simple selection factorization model.  We assume the complete data is a standard normal, i.e. $\theta_Y = (\mu, \sigma) = (0, 1)$, and that the selection mechanism logistic, linear in the data $y_i$.  Under this data generating process, the observed data density is
\[
 f^{\rm obs}(y_i \mid r_i=1, \theta_{Y}, \theta_{R|Y}) \propto \frac{\exp\left(\frac{(y_i-\mu)^2}{2\sigma^2}\right)}{\sigma(1+\exp(-\beta_0-\beta_1y_i))},
 \]
which cannot 
be represented as a finite mixture of exponential families.

For the analysis, we again assume the goal is to estimate the complete data mean and standard deviation.  
We use a mixture of normals to approximate the observed data density but take the parameters of the logistic missingness mechanism, $\theta_{R|Y} = (\beta_0, \beta_1)$ to be known exactly. 
We estimate the posterior distributions for $\mu$ and $\sigma$ for increasing sample sizes and for different values of the selection parameters, $\beta_1$.  
More in detail, in each simulation $\beta_0$ is fixed to 1/2, the parameter $\beta_1$ takes values 1, 2 and 5, and we  replicate the analysis for 100 and 1,000 data points.
To increase the realism of the data analysis, we fit a mixture model with 3 components to 100 data points, but we increase the flexibility of the model for the larger sample size by fitting  a mixture model with 5 components to 1,000 data points.
\begin{figure}[b!]
\centering    
\subfigure[Complete Data Mean]{
 \label{fig:robust-mean}
 \includegraphics[width=0.6\textwidth]{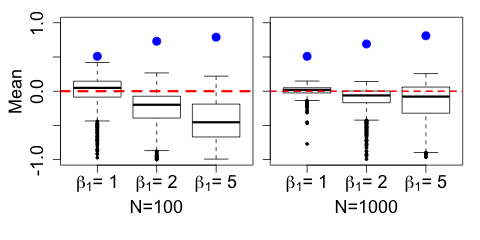}} \\
\subfigure[Complete Data Standard Deviation]{
 \label{fig:robust-sd}
 \includegraphics[width=0.6\textwidth]{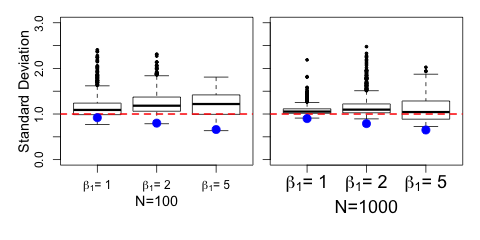}} \caption{\onehalfspacing Posterior estimates of the complete data mean and complete data standard deviation as a function of  sample size and the slope of the missingness mechanism.  Blue dots represent the empirical observed data estimates and the dashed red line corresponds to the true complete data quantity.  As the slope, $\beta_1$ increases, the observed data density approaches a truncated normal, for which Tukey's representation is not applicable.}
\label{fig:robustness}
\end{figure}

Figure \ref{fig:robustness} shows box plots of the posterior distributions for the complete data mean (top panels) and complete data standard deviation (bottom panels) for the different values of $\beta_1$ and increasing sample sizes.  To better appreciate the results, note that, as $\beta_1$ increases, the selection probability approaches zero for all $y_i < \beta_0$ and the observed data density approaches a truncated normal model. Because of the positivity condition, such a model cannot be estimated using Tukey's representation.  The results  in Figure \ref{fig:robustness} show that even when the positivity condition holds, in theory, when the selection probabilities are very close to zero, the complete data estimates (blue dots)  become unstable, increasing the bias of the complete data estimates. Thus, in this example, increasing sample size only partially offsets the failure of inference procedures based only on the observed data.

More generally, instability occurs when the odds of selection are  small in regions where the inferred observed data density is non-zero.  This insight can be formalized  by examining the interaction of the observed data density and the selection probabilities in the Equation \ref{miss_dist}.  As a consequence, the fit of the observed data distribution in the tails of the distribution can be very important for accurately inferring complete data quantities.  In practice, caution is needed when using Tukey's representation in cases where the observed data sample size is small or where the selection probabilities diverge rapidly.

\subsection{Analysis of transcriptomic and proteomic data}
\label{sec:app}

In experiments involving measurements of transcriptomic and proteomic
data, mRNA transcripts and proteins which occur at low levels are less
likely to be observed \citep{Walther2010,Soon2013}.  This makes it challenging to infer
normalizing constants for absolute protein levels
\citep{karpievitch12}, cluster genes into functionally related sets
\citep{olga01}, infer the degree of coordination between transcription
and translation \citep{franks2014}, and determine the ratio of dynamic
range inflation from transcript to protein levels \citep{Csardi2015}.
Here, we demonstrate how data analysis with Tukey's representation can be used to
investigate some these issues by assessing the sensitivity of
estimands to different assumptions about the missingness mechanism.  

In
this analysis, we use transcriptomic data from \citet{pelechano10}, with 14\% missing data, and protein abundance data from \citet{ghaem03}, with 34\% missing data. For illustrative purposes we treat the complete
data mean and variance as the estimands of interest in the analysis.

It is standard to assume that both mRNA and protein levels are
log-normally distributed \citep{beng2005,lu2007}, although this
assumption may not be justified \citep{Lu09,marko2012}.  Here, we
instead model the observed data as a mixture of normals and specify a
prior distribution for the parameters of the logistic missingness mechanism.  Together these
assumptions imply a more flexible prior distribution over complete data densities.

Further, as noted in \citet{Karpievitch2009}, missing values can occur for
multiple reasons, at different stages of the data collection process.
They find that a small fraction of missing proteomic
data collected using mass spectrometry is missing completely at
random.  Consistent with this, we again generalize the results of Section
\ref{sec:ef} to allow the selection mechanism to have a logistic
form that asymptotes at some value less than one.  The observed data
distribution and missingness mechanism together define the joint
distribution:
\begin{align}
\label{app:obs}
f^{\rm obs}(y_i | r_i=1,\beta,\kappa,\mu_k,\sigma_k) &\sim \sum_{k=1}^K
                                         w_kN(y_i;\mu_k,\sigma_k^2)\\
f(r_i=1|y_i,\theta_{R|Y}) &= \frac{\kappa e^{\beta_1y_i+\beta_0}}{1+e^{\beta_1y_i+\beta_0}}
\end{align}
with $\theta_{Y|R=1} = (\mu,\sigma,w)$ and
$\theta_{R|Y} = (\beta_0,\beta_1,\kappa)$.  Here, $0 < (1-\kappa) < 1$
corresponds the fraction of data that is missing completely at random, and
$\beta_0$ and $\beta_1$ describe the odds of a missing value, with
$\beta_1$ parametrizing the rate at which the odds of a missing value
change in $y_i$.  Under this model, the implied missing data
distribution is
\begin{align}
\label{app:miss}
\nonumber f^{\rm miss}(y_i|r_i=0,\beta,\kappa,\mu_k,\sigma_k) &= (1-\kappa^*)
f^{\rm obs}(y_i|r_i=1,\mu_k,\sigma_k) \\
& + \kappa^*\left(\sum_{k=1}^K w_k^*N(y_i;\mu_k+
\beta_1\sigma_k^2,\sigma_k^2)\right)
\end{align}
The full derivation of the mixture weights
$w^*_k$ and $\kappa^*$ is given in Appendix \ref{app:sec:app}.  

\begin{figure}[ht!]
\centering     
 \includegraphics[width=0.425\textwidth]{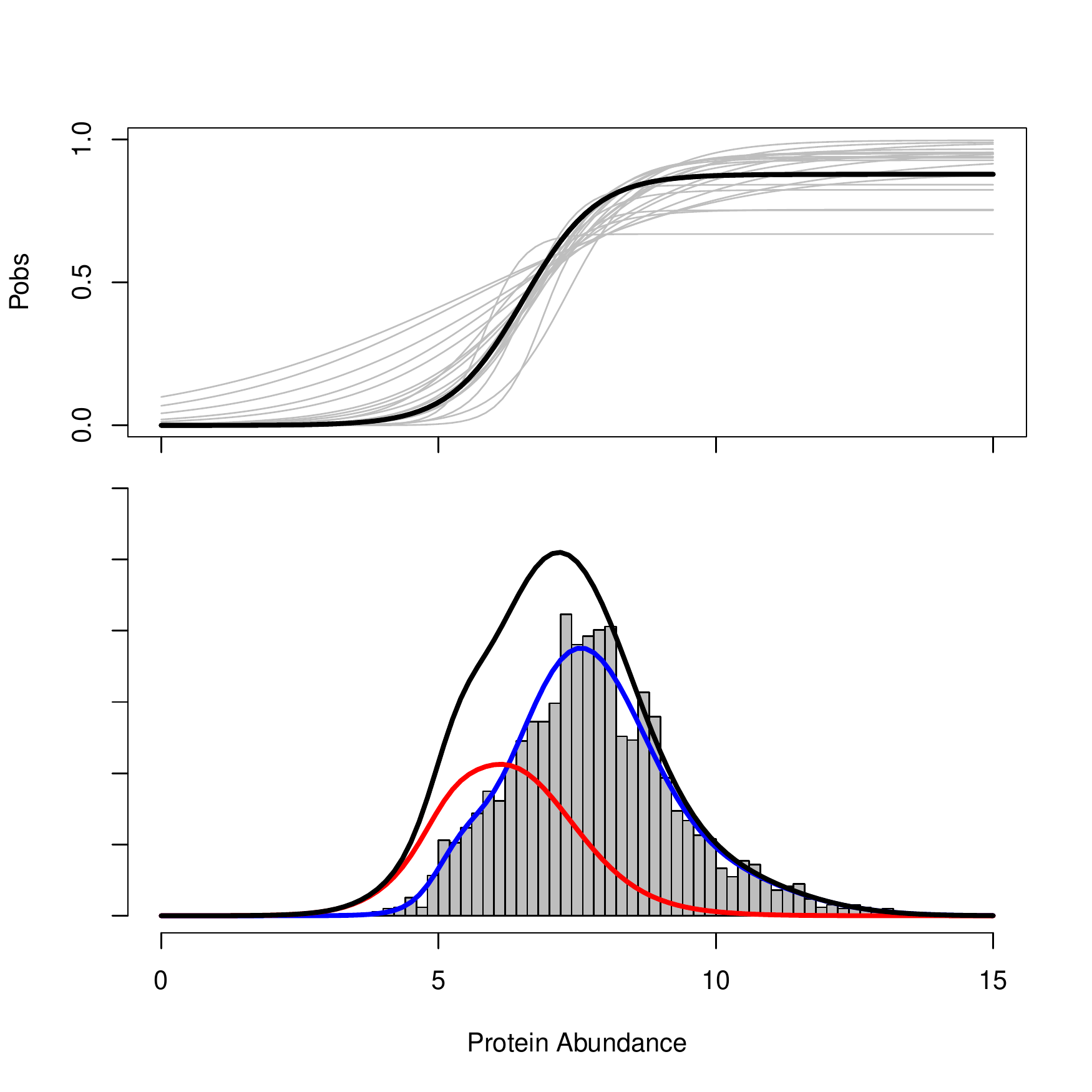}
 \includegraphics[width=0.425\textwidth]{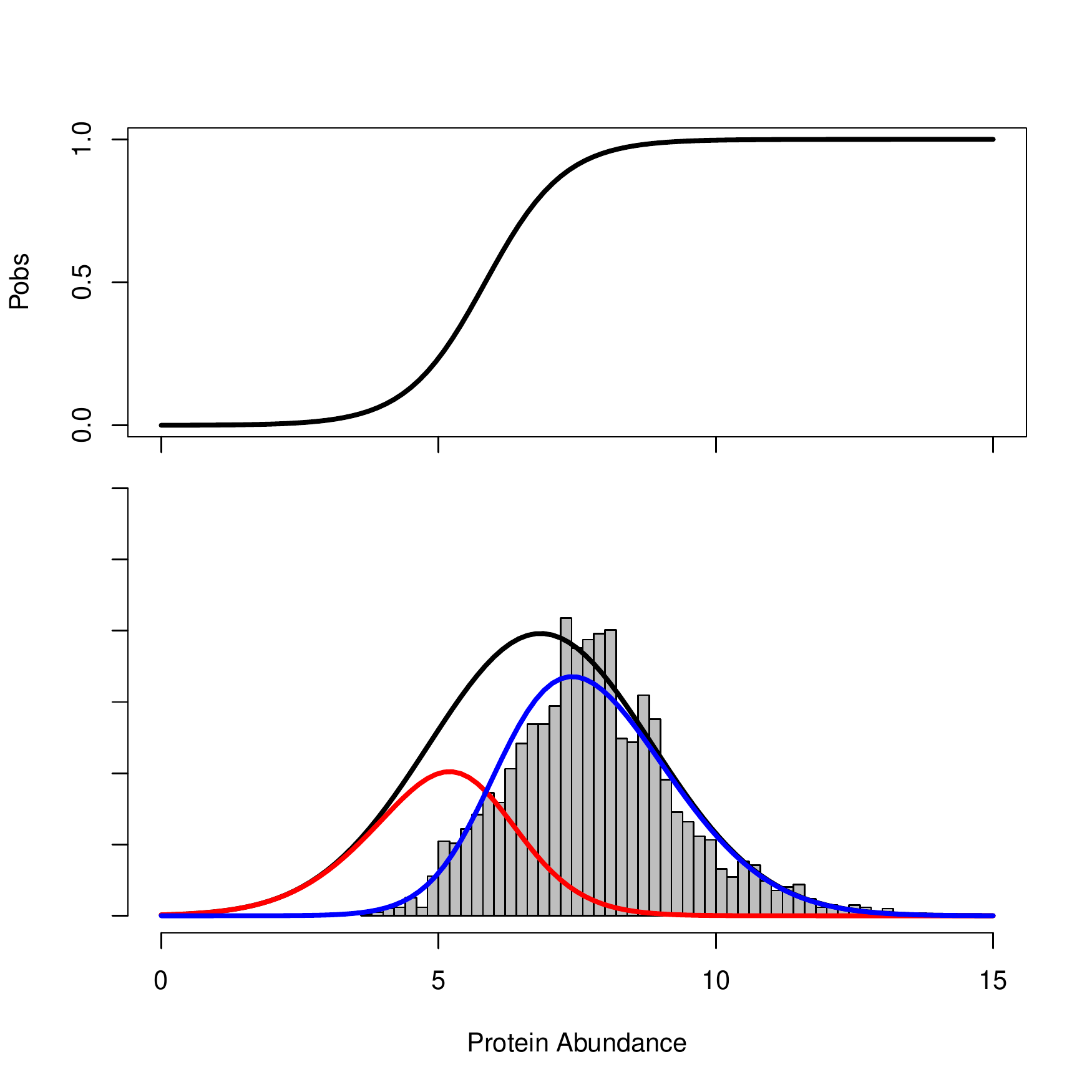}
\caption{\onehalfspacing Model fit to proteomic data from
  \citet{ghaem03} data using two different approaches: Tukey's
  representation (left) and the selection factorization (right).  The
  grey lines in the top-left panel represent draws of the selection
  mechanism from the prior distribution provided in Equation \ref{app:selprior}. The black, red and blue lines in the bottom panels correspond to the estimated densities of complete data, missing data and observed data, respectively.}
\label{fig:seldist2}
\end{figure}

In the analysis of the observations, to poist a good model for $f^{\rm obs}$ in  Tukey's representation, we
found that $K=3$ components were enough to approximate the
observed data well.  Furthermore, we chose the following prior
distributions for Q and $\theta_{R|Y}$,
\begin{eqnarray}
\label{app:selprior}
 \beta_1  &\sim&  {\rm Beta}(1,3) \nonumber\\
  Q       &\sim&  {\rm Uniform}(0,1) \nonumber\\
 \kappa   &\sim&  1-(1-Q){\rm Beta}(2,1), \quad \kappa \geq Q
\end{eqnarray}
Note that $\kappa$ must be greater than $Q$ because the selection
probabilities cannot be  less than the population fraction
of observed data. Draws from this prior distribution are shown in Figure \ref{fig:seldist2}, top-left panel, in grey, around the prior mean, in black. We
implemented the sampler for the making inference with this model
using  STAN \citep{stan}.

\begin{figure}[ht!]
\centering   
 \includegraphics[width=0.425\textwidth]{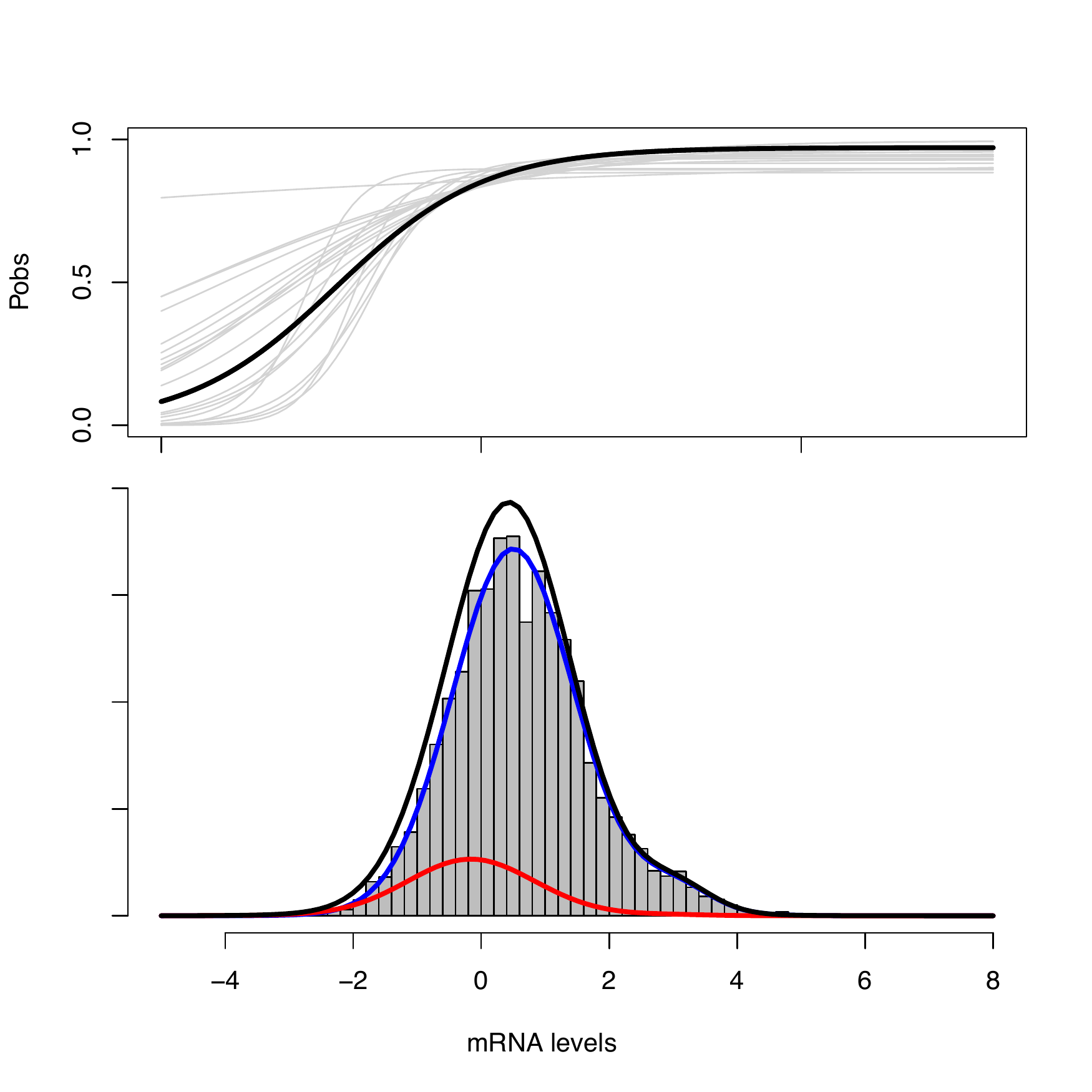}
 \includegraphics[width=0.425\textwidth]{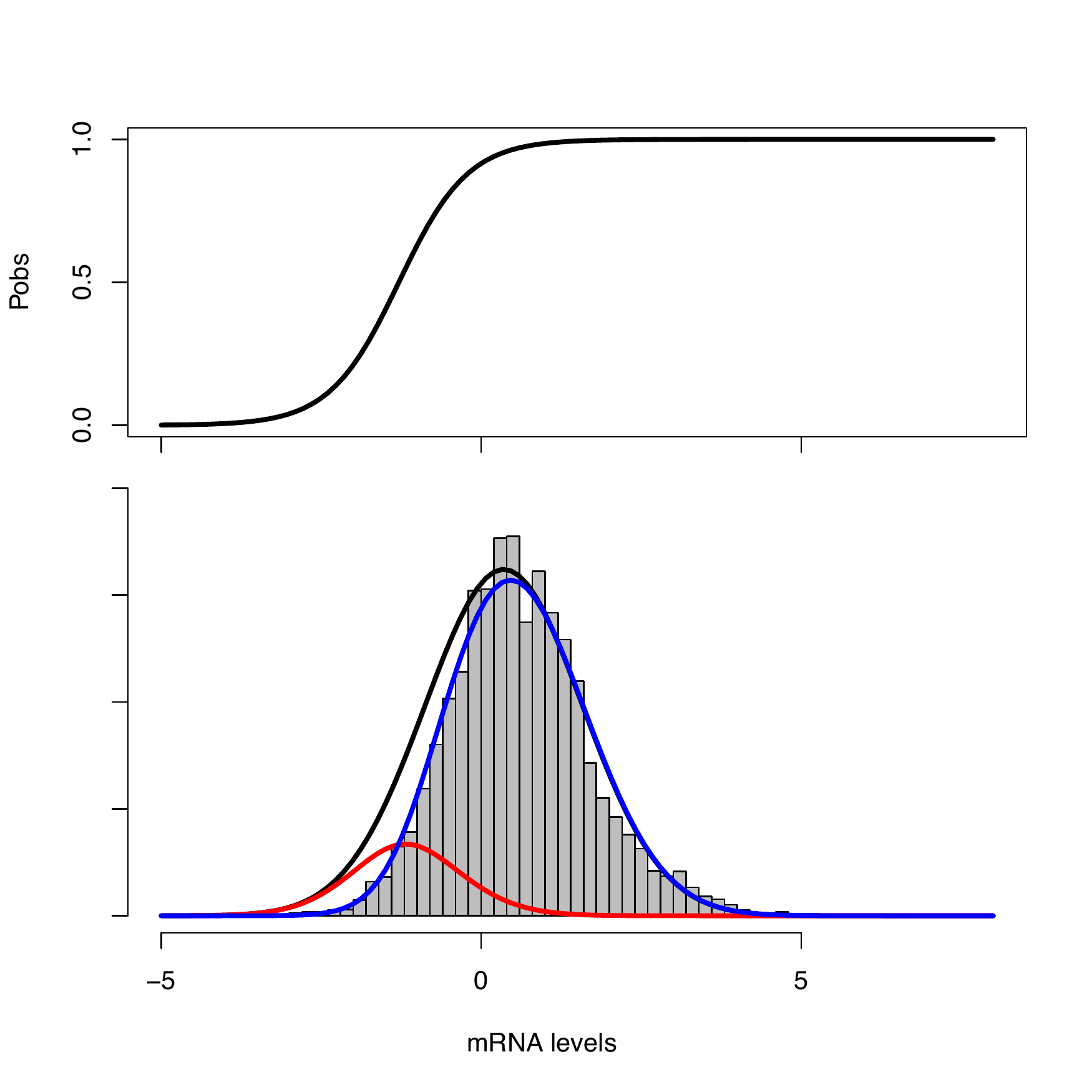}
\caption{\onehalfspacing Model fit to mRNA data from
  \citet{pelechano10} data using two different approaches: Tukey's
  representation (left) and the selection factorization (right). The
  grey lines in the top-left panel represent draws of the selection
  mechanism from the prior distribution provided in Equation \ref{app:selprior}. The black, red and blue lines in the bottom panels correspond to the estimated densities of complete data, missing data and observed data, respectively.}
\label{fig:seldist1}
\end{figure}

Figure \ref{fig:seldist2}, bottom-left panel, shows the fit to the protein data \citep{ghaem03}, when $\beta_1$ is set to its median posterior value.
For comparison, the bottom-right panel, shows the fit of the selection factorization model published in \citet{franks2014}, which  assumes the complete data  is distributed according to a lognormal and the missingness mechanism is logistic, with the mean linear in in $y_i$. 
The black, red and blue lines, in both  bottom panels, correspond to the estimated densities of complete data, missing data and observed data, respectively.
Lack-of-fit to the observed data is the analysis with the selection model.  
Figure \ref{fig:seldist1} shows the corresponding results for the transcript data set \citep{pelechano10}, where the lack-of-fit using the selection factorization is less pronounced.


\begin{figure}[t!]
     \centering
     \includegraphics[width=0.85\textwidth]{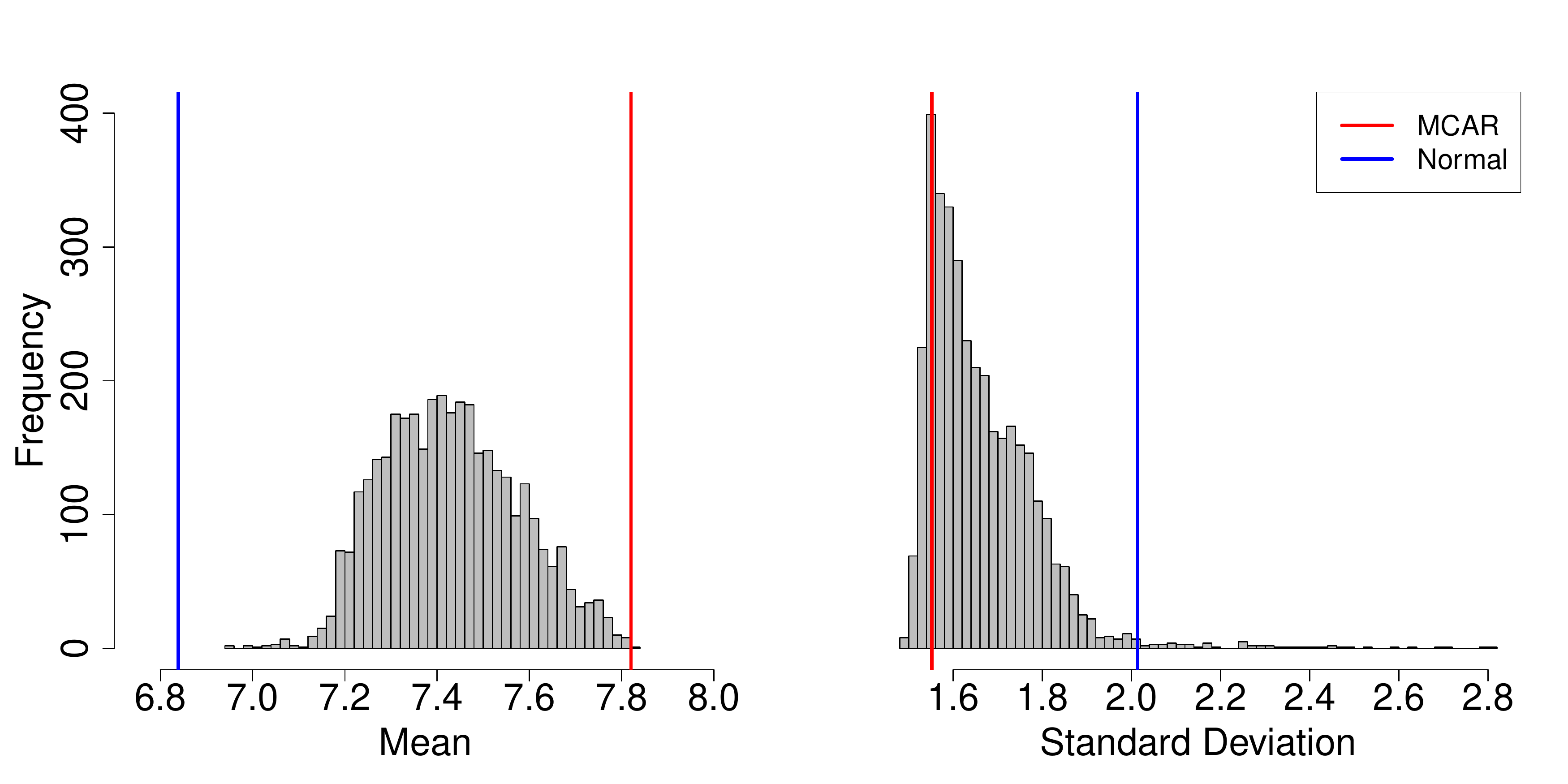}
     \caption{\onehalfspacing Posterior distributions of the complete data mean (right) and
       complete data standard deviation (left) for protein data
       \citep{ghaem03}.  The MCAR estimates (red) and an estimate
       assuming normality of the complete data (blue) are shown as
       vertical lines for comparison.  Under the prior distribution in Equation \ref{app:selprior}, estimates using the MCAR
       and the selection factorization models are at opposite ends of these posterior distributions. }
    \label{fig:comparisons2}
\end{figure}

\begin{figure}[t!]
     \centering
     \includegraphics[width=0.85\textwidth]{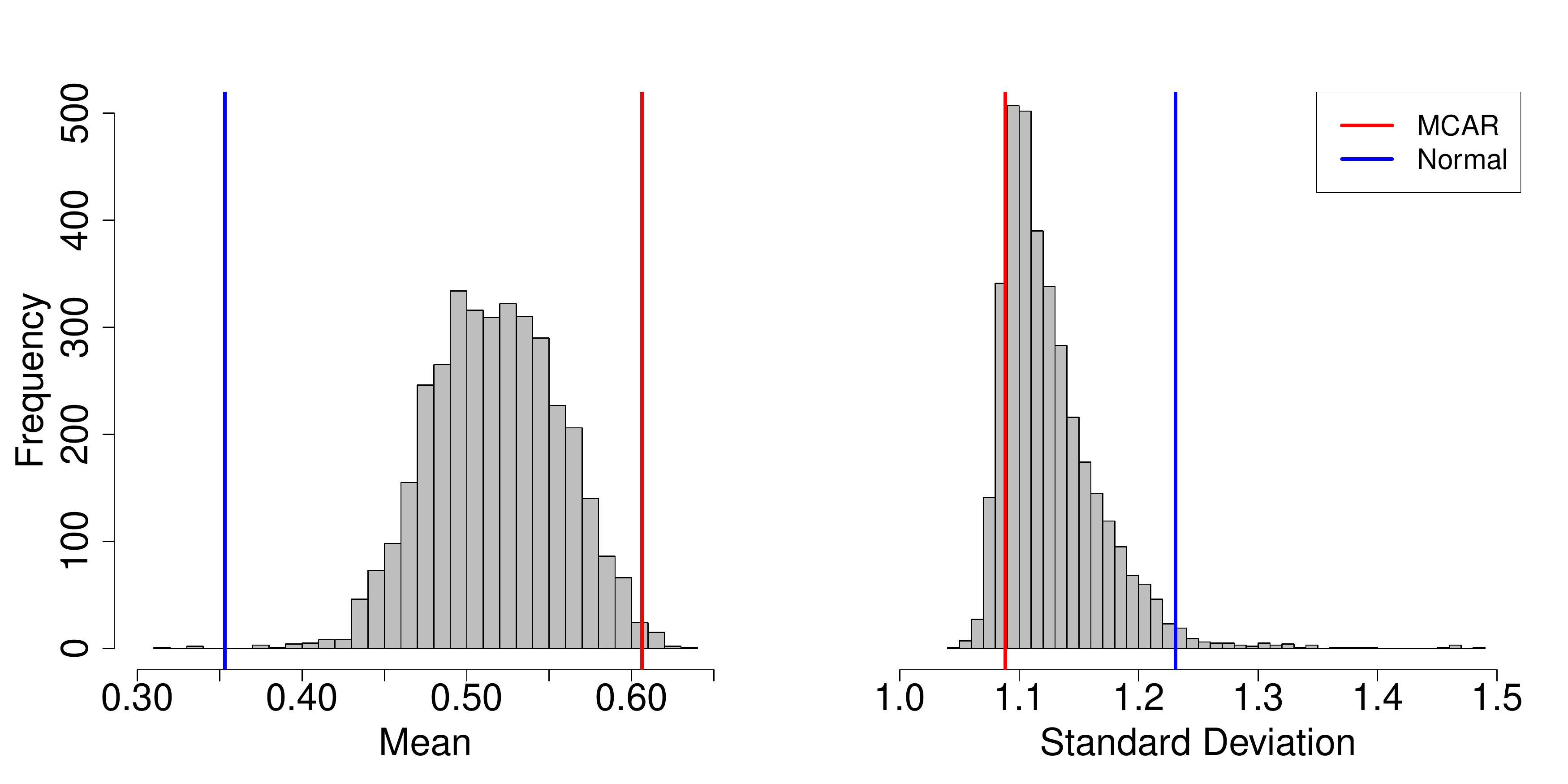}
     \caption{\onehalfspacing Posterior distributions of the complete data mean (right) and
       complete data standard deviation (left) for mRNA data
       \citep{pelechano10}.  The MCAR estimates (red) and an estimate
       assuming normality of the complete data (blue) are shown as
       vertical lines for comparison.  Under the prior distribution in Equation \ref{app:selprior}, estimates using the MCAR
       and the selection factorization models are at opposite ends of these posterior distributions. }
    \label{fig:comparisons1}
\end{figure}

In Figures \ref{fig:comparisons2}--\ref{fig:comparisons1}, we compare the estimated complete data posterior means and standard deviations  for the protein and transcript data sets, respectively, using three different missing data models: 
 Tukey's representation model implied by Equations \ref{app:obs} and \ref{app:selprior};
 a selection factorization model that assumes log-normality of the complete data and a logistic selection mechanism \citep{franks2014,Csardi2015}; and
 a missing complete at random model.
For both data sets, the estimates obtained with  the MCAR model and with the selection factorization models bookend the estimates obtained with  the model specified with Tukey's representation. 
 Under the selection
factorization model, the complete data standard deviation is large and
the mean is small relative to the estimates from Tukey's
factorization.  This is likely because the strong parametric assumptions
associated with the  selection factorization models overly constrain the fit
to the observed data.  
%
 Table \ref{app:table} reports exact numerical estimates of the two estimands of interest, using the three competing models.

Recent published analyses of data using the selection factorization found that translational regulation widens the dynamic range of protein expression \citep{franks2014,Csardi2015}. One way to quantify the relative dynamic ranges of mRNA and protein is by computing the ratio of the standard deviations between log-mRNA and log-protein levels, a protein-specific quantity often referred to as the ``dynamic range ratio''.  A value of this ratio less then one  suggests that the dynamic range of protein levels is smaller than that of mRNA, and is taken as evidence of a suppressive role of translational regulation.  A value greater than one is  taken as evidence of amplification, as claimed by the recent analyses.

\begin{figure}[t!]
     \centering
     \includegraphics[width=0.5\textwidth]{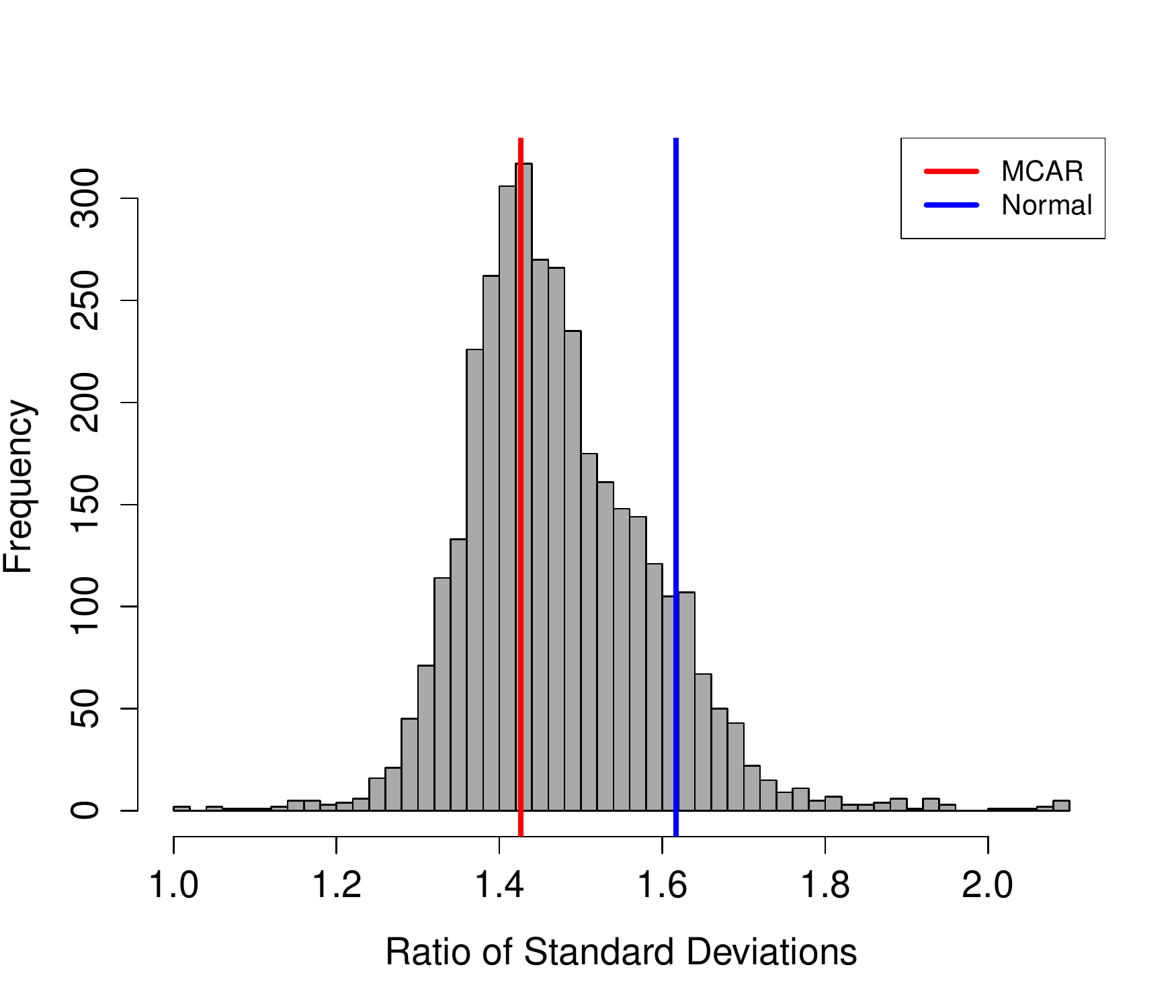}
     \caption{\onehalfspacing Dynamic range ratios obtained using Tukey's representation (histogram), the selection factorization model  (blue) of \citet{Csardi2015}, and an MCAR model (red).}
    \label{fig:ratio}
\end{figure}

We used posterior estimates of the complete data standard deviations, obtained from the three competing models fit to both protein and mRNA data sets \citep{pelechano10,ghaem03} to estimate the distribution of the dynamic range ratios, in Figure \ref{fig:ratio}.  The results obtained with Tukey's representation are consistent with those reported by \citet{Csardi2015}, suggesting that translational regulation reflects amplification of protein levels.
 Table \ref{app:table} also reports  numerical estimates of the the dynamic range ratios.

\begin{table}[t!]
\centering
\begin{tabular}{l|c|c|c|l}
 \belowspace
 Estimand & Tukey's representation & Selection factorization  & MCAR & Data set \\ \hline
 \abovespace
      Mean & 7.42 ~(7.18, 7.73) & 6.84 & 7.82 & Prot. \\
 \belowspace
 Std. Dev. & 1.66 ~(1.52, 1.94) & 2.01 & 1.55 & Prot. \\ \hline
 \abovespace
      Mean & 0.51 ~(0.44, 0.59) & 0.35 & 0.60 & mRNA \\
 \belowspace
 Std. Dev. & 1.13 ~(1.07, 1.23) & 1.23 & 1.08 & mRNA \\ \hline
 \abovespace
    Ratios & 1.48 ~(1.28, 1.73) & 1.62 & 1.43 & Both \\
\end{tabular}
\caption{\onehalfspacing Estimates for the quantities of interest obtained with different models, from protein and mRNA data. The dynamic range ratios are computed using both data sets.  We report  maximum likelihood point estimates for both the MCAR and selection models. We report posterior medians and 95\% posterior intervals (in parentheses) for Tukey's representation.}
\label{app:table}
\end{table}

These results demonstrate the relative ease of applied data analysis  with
Tukey's representation models, and the increased flexibility of models
specified using this full conditional specification.  By directly modeling the observed
data, we avoid the need for Monte Carlo integration of the missing
data, and do not require parametric specifications for the complete data
density as is typical for selection models.  By modeling the selection
function directly, we are also able to express uncertainty about the
missing data density beyond the simple location and scale changes
typical in pattern-mixture model sensitivity analyses.



\section{Discussion}
\label{sec:disc}
Tukey's representation provides a powerful alternative for specifying missing data models.  
It allows analysts to eschew some difficult questions about identifiability in models for non-ignorable missing  data \citep{miao2015} by factoring the joint distribution of the complete-data, $Y$, and missing-data indicators, $R$, in such a way that the missingness mechanism is the only component that must rely on assumptions unassailable using observed data.

\subsection{Theoretical insights}

Thus far we largely worked with exponential-family models. Here, we make formal statements about exponential family, and give results that hold  in greater generality.

\begin{theorem}
The normalizing constant $Q$ is the population fraction of observed data.
\end{theorem}
The proof is straightforward because 
\begin{eqnarray}
\label{q:er}
 \mathbb{E} [ r_i \mid \theta_{Y|R},\theta_{R|Y}]
 &=& f(r_i=1\mid \theta_{Y|R},\theta_{R|Y})  \\
 &=& \int f(y_i, r_i=1\mid \theta_{Y|R},\theta_{R|Y}) ~dy_i \nonumber \\
 &=& Q (\theta_{Y|R},\theta_{R|Y}) ~\int f^{\rm obs}(y_i\mid \theta_{Y|R}) ~dy_i \nonumber\\
 &=& Q \nonumber
\end{eqnarray}

The density of the observed values, $y^{\rm obs}$ and $r$, can be written in terms of the population fraction of the observed data, $Q$, in full generality,
\begin{eqnarray}
\label{lik:obs}
f(y^{\rm obs},r \mid \theta_{R|Y},\theta_{Y|R})
 &=&        \prod_{\{i : r_i=1\}} f^{\rm obs} (y_i \mid \theta_{Y|R}) ~Q \times \nonumber \\
 & & \times \prod_{\{i : r_i=0\}} \int \frac{f(r_i=0\mid y_i, \theta_{R|Y})}{f(r_i=1\mid y_i, \theta_{R|Y})} f^{\rm obs} (y_i \mid \theta_{Y|R}) ~Q ~dy_i \nonumber \\
 &=&        Q^{\sum_i r_i} (1-Q)^{N-\sum_i r_i} \prod_{\{i : r_i=1\}} f^{\rm obs} (y_i \mid \theta_{Y|R}),
\end{eqnarray}
using the observation that $\frac{\int \dots \,dy_i}{1+\int \dots \,dy_i} = 1-(1+\int \dots \,dy_i)^{-1} = 1-Q$.
And the missing-data density can  be expressed as a function of the observed-data density in full generality. 
\begin{theorem}
The missing-data density can be expressed a function of the observed-data density and the odds of missingness,
\begin{equation}
\label{miss_dist}
f^{\rm mis}(y_i \mid \theta_{R|Y}, \theta_{Y|R}) =
 \frac{Q (\theta_{Y|R},\theta_{R|Y})}{1-Q (\theta_{Y|R},\theta_{R|Y})}
 \frac{f(r_i=0\mid y_i, \theta_{R|Y})}{f(r_i=1\mid y_i, \theta_{R|Y})}
 ~f^{\rm obs} (y_i \mid \theta_{Y|R}).
\end{equation}
\end{theorem}
This result 
can  be derived from from the complete-data likelihood and the formula for $Q$ in Equations \ref{lik:comp} and \ref{eqn:normconst}.  

When the missingness mechanism, $f(r_i \mid y_i, \theta_{R|Y})$ does not depend on $y_i$, the distribution of the missing data is directly proportional to the distribution of the observed data. 
Equation \ref{miss_dist} can help assess the plausibility of different missingness mechanisms---{\em not at random}, {\em completely at random}, and {\em at random}  \citep{Mealli:2015aa}---by  viewing them  as functions of the odds of a missing value, $\frac{f(r_i=0\mid y_i, \theta_{R|Y})}{f(r_i=1\mid y_i, \theta_{R|Y})}$.  For instance, when the odds  have low variance, it may be reasonable to assume the missing data mechanism is completely at random, or at random.

Equation \ref{miss_dist} also leads to a general understanding of the main result of Section \ref{sec:ef}, which can be crystallized in the following statement. 
\begin{theorem}
If the observed-data distribution, $f^{\rm obs}$, belongs to an exponential family, and the log-odds of a missing value are linear in the natural sufficient statistics of that observed-data distribution, then the  missing-data distribution, $f^{\rm mis}$, must have the same exponential family as the observed data distribution.
\end{theorem}
This result can be immediately extended to mixtures.
\begin{corollary}
If the observed-data distribution, $f^{\rm obs}$, can be expressed as a $K$-component mixture of an exponential family distribution, and the log-odds of a missing value are linear in the natural sufficient statistics of that observed-data distribution, then the implied missing-data distribution, $f^{\rm mis}$, must be a $K$-component mixture of the same exponential family.
\end{corollary}
We conjecture that a similar relation might hold outside exponential family models.

\subsection{A note on the integrability condition} 
\label{sec:integrability}


Not all integrable specifications for
$f^{\rm obs}(y_i \mid\theta_{Y|R=1})$ and $f(r_i\mid y_i,\theta_{R|Y})$ imply a
proper  distribution for $f^{\rm mis}(y_i \mid \theta_{Y|R=1},\theta_{R|Y})$.  In the setting we consider in Section \ref{sec:bayesian}, the integrability condition requires the sum $\theta_{Y|R=1}+\theta_{R|Y}$ to lie in the natural parameter space of the exponential family. 
In practice, analysts may want to consider  missing data mechanisms that involve a richer set of parameters, $\tilde\theta_{R|Y}$, such as including an intercept, as we illustrate in the context of a biology application, in Section \ref{sec:app}. In such cases, $\theta_{R|Y}$ is taken to denote the subset of parameters in $\tilde\theta_{R|Y}$ that multiply the sufficient statistics of $f^{\rm obs}$.
The derivations in Section \ref{sec:bayesian} are simple to repurpose for this situation, accordingly. 

For example, assume that the natural parameter $\theta_{Y|R=1} =
\eta$, and that the missing data mechanism is logistic with extended parameter vector $\tilde\theta_{R|Y}
= (\beta_0,\beta_1) = (\beta_0,\theta_{R|Y})$ and $f(r_i = 1 \mid y_i,\beta) =
(1+e^{-(\beta_0+T(y_i) \beta_1)})^{-1}$. Then, Equations \ref{eqn:exfam} and \ref{miss_dist_ef} become
\begin{align}
\label{eqn:intercept}
Q(\eta,\beta) &=
  \frac{g(\eta+\beta_1)}{g(\eta+\beta_1)+g(\eta)e^{\beta_0}}\\
f^{\rm mis}(y \mid \eta,\beta) &= h(y) g(\eta + \beta_1) e^{{T(y)}'(\eta+\beta_1)}.
\end{align}

The class of exponential-logistic models defined in Equations \ref{eq:exp-log-obs}--\ref{eq:exp-log-sel} can be further generalized in two useful ways while maintaining its desirable properties.
For instance, generalizing $f^{\rm obs}$ to be a mixture of
exponential families as in Section \ref{sec:sim} is straightforward, and does not increase
computation substantially.  Relaxing assumptions about the missingness
mechanism can be more difficult.  Still, it is possible to model
$f(r_i \mid y_i,\theta_{R|Y})$ with a mixture of logistic functions, including a  missingness mechanism  where a fraction of the data is missing completely at random as in Section \ref{sec:app}.

\subsection{Inference strategies}

Recall the simple Normal-logistic model of Section \ref{sec:est:inf},
\begin{eqnarray}
 f(r_i = 1 \mid y_i,\beta) &=& \hbox{logit }(\beta_0+\beta_1 y_i) 
                            = \bigm(1+\exp\{- \beta_0-\beta_1 y_i\}\bigm)^{-1} \nonumber \\
 f^{\rm obs}(y_i)          &=& \hbox{Normal }(0,1). \nonumber
\end{eqnarray}
The inferential strategy  proposed was to posit prior distributions on $\beta_1$ and $Q$, and solve for $\beta_0$ at each iteration of the Markov Chain Monte Carlo sampler.
A conceptually simpler approach to  inference would be to place a prior distribution on all the parameters of the missingness mechanism, and solve for the implied $Q$ at each iteration of the sampler.

In situations where the number of missing values is itself missing, as with truncated data, specifying a   prior distribution for all the parameters of the missingness mechanism would lead to an implied prior distribution for the unknown number of missing values, or equivalently, the population fraction of observed data Q. 
\begin{figure}[b!]
 \centering
  \includegraphics[width=78mm]{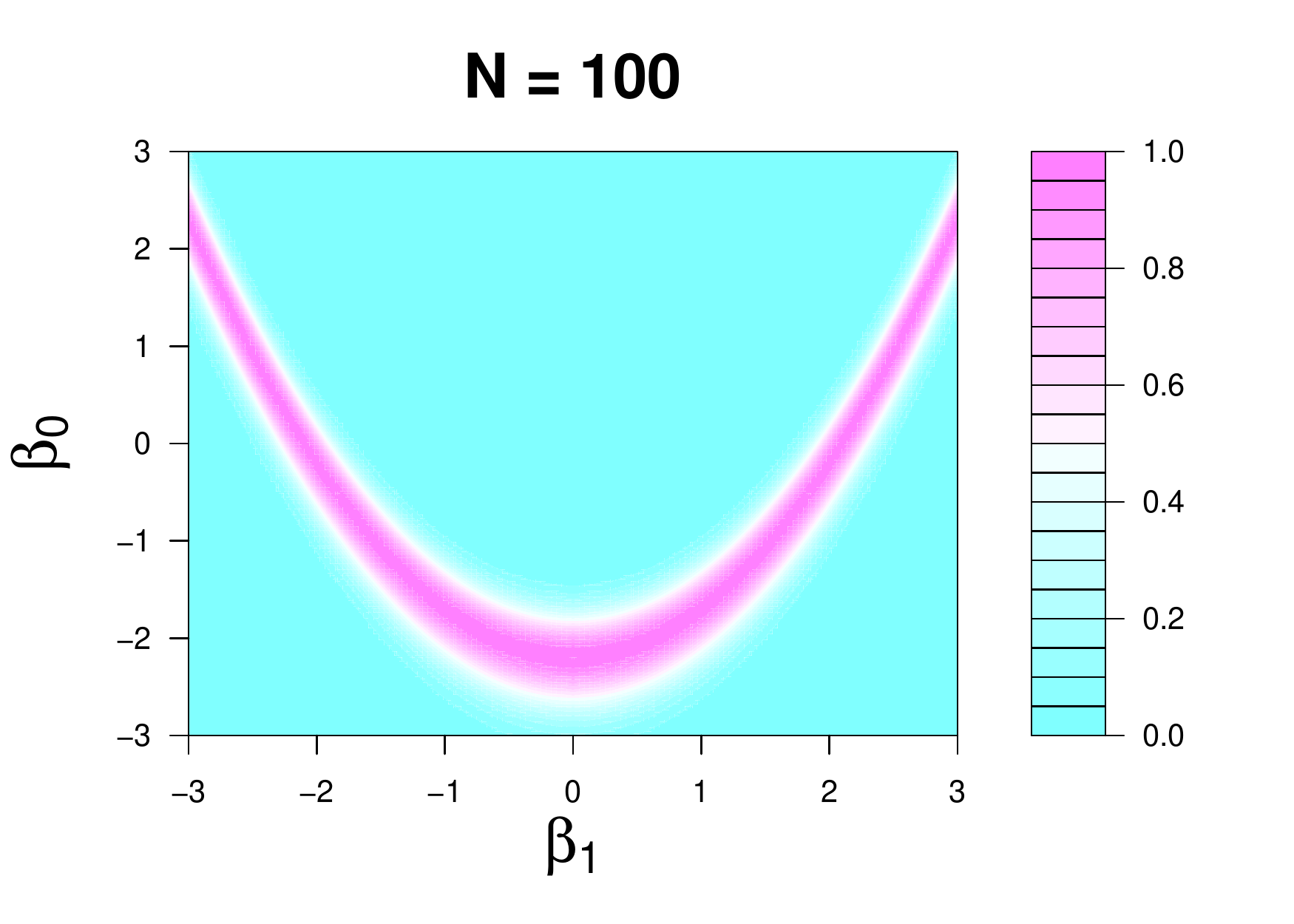}
  \includegraphics[width=78mm]{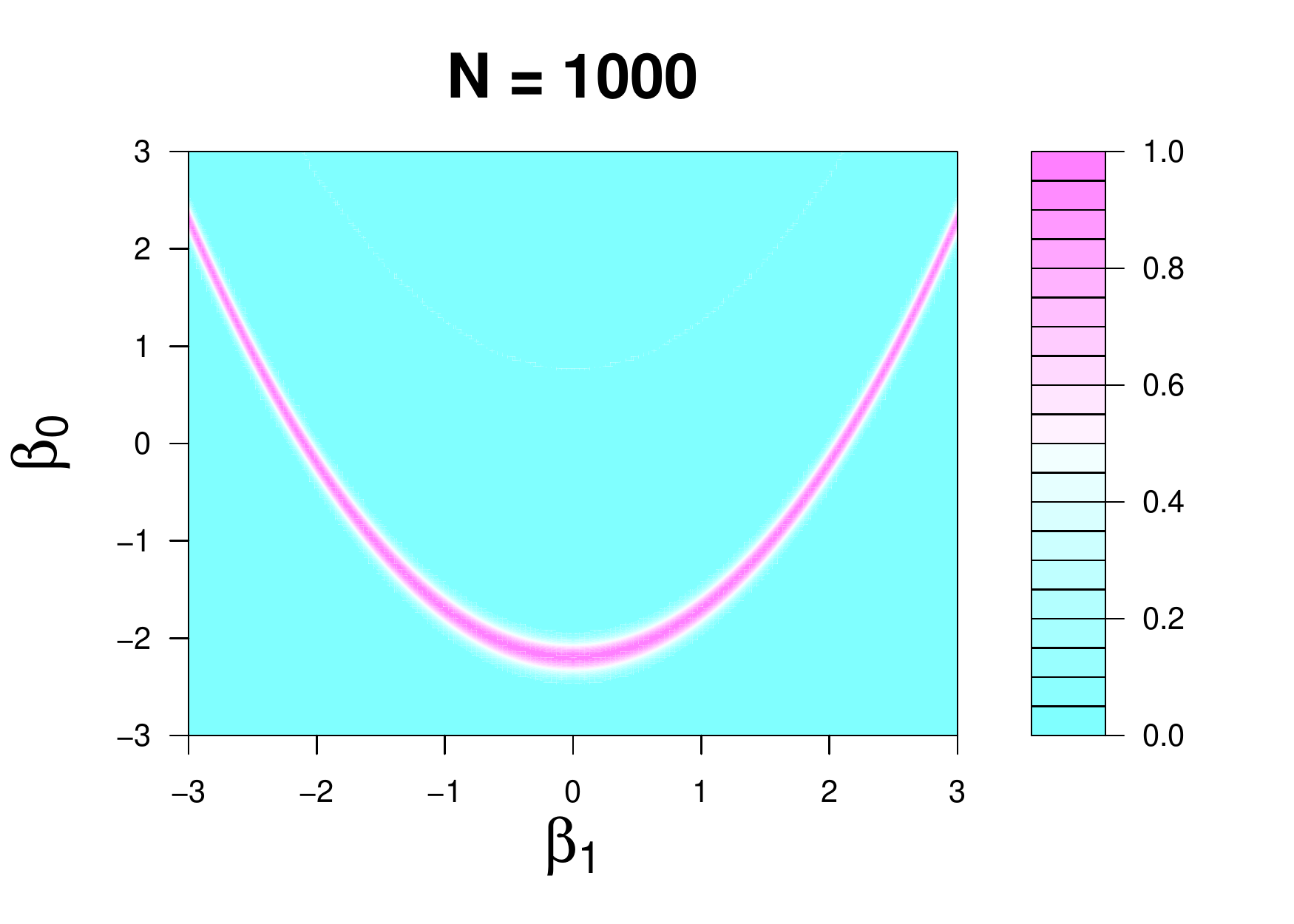}
\caption{\onehalfspacing The region of positive support for the likelihood, restricted to the parameters of the missingness mechanism, is increasingly constrained as
  the population fraction of missing data $Q$ is estimated  with
  increasingly high precision. This intuition is illustrated by the
  width of the ridge, which is a function of the amount of information
  about the fraction of missing data $Q$. We simulated data from a
  standard Normal distribution and a logistic missingness mechanism. The parameters $(\beta_0,\beta_1)$ were set to get 90 percent missing data. The sample size determines the amount of information: $N=100$ (left) and $N=1000$ (right).}
    \label{fig:contour1}
\end{figure}

In situations where the number of missing values is known, however, as with censored data, and therefore $Q$ can be estimated from observed data, the support of the likelihood is a constrained parameter space, and a number of choices for the prior distribution on $\beta = \theta_{R|Y}$ would lead to a posterior distribution that is challenging to explore using Monte Carlo  methods.
Specifically, Equation \ref{q:er} induces a moment constraint that restricts the region where the parameters of the missingness mechanism have positive support to a lower dimensional ridge.
Figure \ref{fig:contour1} illustrates this phenomenon for the simple Normal-logistic model, and increasing sample size.

Sequential Monte Carlo and other specialized Monte Carlo methods that  exploit the geometry of the support of posterior distribution may provide a  solution in this situation \citep{DFG2001aa, liu2008aa, Girolami2011, airohaas:2010}.

\subsection{Concluding remarks}

In
this paper, we used  logistic-exponential family models
to illustrate how Tukey's representation can be used to encode non-monotonicity in the
missingness mechanism, 
and to model data with complex
distributional forms.  These models could also be used to facilitate tipping
point analyses \citep{Liublinska2014}, or to incorporate
subjective model uncertainty via prior distributions on the missingness mechanism \citep{rubi:2004}.

Tukey's representation is most useful when positing
reasonable prior distributions on the selection mechanism is feasible.  Translating expert
knowledge or expectations into a functional form can be
challenging, in general, and a logistic missingness mechanism is
not always a good choice.  In practice, Tukey's representation should be used
in concert with strategies for expert prior elicitation
\citep{ohagan06,kynn05,paddock09}. 
Nevertheless,  prior elicitation for Tukey's representation is simpler than for other factorizations, since it involves only the  set of parameters $\theta_{R|Y}$.
In contrast, the selection factorization requires  additional assumptions about the complete data density.

In many settings, like the example presented in Section
\ref{sec:app}, we may be able to collect data which partially informs
the specification for the selection mechanism.  As such, when
possible, we can design experiments to learn about the functional form
of $f(r_i|y_i,\theta_{R|Y})$ as well as to further refine prior distributions for $\theta_{R|Y}$.
Along these lines, Tukey's representation may be useful in the context
of multiphase inference, which is intimately related to problems in
missing data \citep{blocker2013}.  In these problems, when
preprocessing data, it is often the case that we have strong knowledge
(or control) of the missingness mechanism yet a weaker understanding of
the underlying scientific model. 
 
All in all, we argue that Tukey's representation, which
represents a hybrid of the selection and pattern mixture models is an
under-researched yet promising alternative for modeling non-ignorable
missing data.

\bibliographystyle{asa}
\bibliography{refs.bib}

\newpage
\appendix
\section{Detailed derivations}
\label{der:sim}


\subsection{Derivations From Section \ref{sec:sim}}
\label{app:sec:sim}

First, assume the observed data are normally
distributed.  For the $k$-th normal component expressed in exponential
family form, with natural parameters $\theta_{Y|R=1}$, we have:
\begin{align}
\theta_{Y|R=1} &= (\eta_1,\eta_2) = (\frac{\mu_k}{\sigma_k^2}, -\frac{1}{2\sigma_k^2})\\
T(y_i) &= (y_i,y_i^2)\\
g(\theta_{Y|R=1}) &=
 \frac{1}{\sqrt{-2\eta_2}}e^{\frac{\eta_1^2}{4\eta_2^2}}
\end{align}
where $\vec \eta$ are the natural parameters, $\vec T(y)$ the
sufficient statistics and $g(\vec \eta)$ is the partition function. By
Equation \ref{sim:mech} the odds of a missing value are
\begin{equation}
\frac{f(r_i=0|y_i,\theta_{R|Y})}{f(r_i=1|y,\theta_{R|Y})} = \beta_2y_i^2-2\beta_2\beta_1y_i+\beta_2\beta_1^2+\beta_0
\end{equation}
with $\theta_{R|Y} = (\beta_0,\beta_1,\beta_2)$.  Finally, by Equation
\ref{miss_dist_ef} the implied missing data distribution for the
$k$-th component is normal
with natural parameters
\begin{equation}
\label{simInverseNat}
(\eta_1^*,\eta_2^*) = (\eta_1-2\beta_2\beta_1,\eta_2+\beta_2)
\end{equation}
The inverse parameter mapping for the normal specifies that
\begin{equation}
 (\mu_k,\sigma_k^2) = \left(\frac{-\eta_1}{2\eta_2},\frac{-1}{2\eta_2}\right) 
\end{equation}
and thus the moments of the missing data distribution are
\begin{align}
\label{sim:mis_moms}
 (\mu_k^*,\sigma_k^{*2}) &=
                      \left(\frac{\frac{-\mu_k}{\sigma_k^2}+2\beta_2\beta_1}{2(\frac{-1}{2\sigma_k^2}+\beta_2)},\frac{-1}{2(\frac{-1}{2\sigma^2}+\beta_2)}\right) \\
\nonumber &=
  \left(\frac{\mu_k - 2\beta_1\beta_2\sigma_k^2}{1-2\beta_2\sigma_k^2},\frac{\sigma_k^2}{1-2\beta_2\sigma_k^2})\right)
\end{align}

We now extend the results of Section \ref{sec:ef} to mixture models.
First,  by Equation \ref{miss_dist_ef},  the observed data
distribution is a mixture of exponential families, the missing data
distribution is also a mixture of those same exponential families.
By applying Equation \ref{eqn:exfam} to a mixture of exponential families
(specifically a mixture of normal and discrete distributions), we have
\begin{align}
Q(w,\eta,\beta) &= \frac{1}{1+\int \left[\alpha\sum_k^K
                  w_kh(y)g(\eta_k)e^{\eta_kT(y)}+(1-\alpha)\sum_j^me^{log p_j}\right](e^{(\beta_2y^2-2\beta_2\beta_1y+\beta_2\beta_1^2+\beta_0)})} \nonumber\\
&= \frac{1}{1+e^{\beta_0+\beta_2\beta_1^2}\left[\alpha\sum_k^K w_k\frac{g(\eta_k)}{g(\eta_k^*)}+(1-\alpha)\sum_j^m h(p_j,\gamma_j,\beta)\right]}
\end{align}
We  invert this equation to express $\beta_0$ as a function of  $Q$, $\eta$ and $(\beta_1,\beta_2)$:
\begin{align}
\label{betaZeroSim}
\beta_0(Q,w,\eta,\beta_1,\beta_2) &=
                                    log\left(\frac{1-Q}{Q\left[\alpha\sum_k^K
                                    w_k\frac{g(\eta_k)}{g(\eta_k^*)}+(1-\alpha)\sum_j^m
                                    h(p_j,\gamma_j,\beta)\right]}\right)
                                    - \beta_2\beta_1^2
\end{align}
with
$h(p_j,\gamma_j,\vec \beta) = e^{\text{log} p_j +
  \beta_2(\gamma_j-\beta_1)^2}$.
Using Equation \ref{miss_dist}, we can show that the mixture
weights are simply
\begin{align}
\label{sim:weights}
w_k^* &= \frac{w_k\frac{g(\eta_k)}{g(\eta_k^*)}}{\sum_k^K
        w_k\frac{g(\eta_k)}{g(\eta_k^*)}}\\
\nonumber \alpha^* &= \frac{\alpha \sum_k^K
  w_k\frac{g(\eta_k)}{g(\eta_k^*)}}{\alpha \sum_k^K
        w_k\frac{g(\eta_k)}{g(\eta_k^*)} + (1-\alpha)
  \sum_j^m h(p_j,\gamma_j,\beta)}
\end{align}

Together, Equations \ref{sim:weights} , and \ref{sim:mis_moms} yield
the missing data distribution specified in \ref{sim:miss}.

\subsection{Derivations From Section \ref{sec:app}}
\label{app:sec:app}

Assume that the observed data can be
represented as a mixture of normals, and also  that the odds of a
missing value can also be represented by a mixture.  Specifically, we
allow for some missing values to occur completely at random.  We
posit that,
\begin{align}
f(r_i=1|y,\theta_{R|Y}) &= \frac{\kappa e^{\beta_1y+\beta_0}}{1+e^{\beta_1y+\beta_0}}
\end{align}
The mechanism is logistic, but asymptotes at some value $\kappa < 1$,
where $(1-\kappa)$ represents the fraction of the complete data that is missing
completely at random.  Under this model, the odds of a missing value are
\begin{align}
\frac{f(r_i=0|y,\theta_{R|Y})}{f(r_i=1|y,\theta_{R|Y})} &= \left(1-\frac{\kappa e^{\beta_1y+\beta_0}}{1+e^{\beta_1y+\beta_0}}\right) \frac{1+e^{\beta_1y+\beta_0}}{\kappa e^{\beta_1y+\beta_0}} \\
    &= \frac{1+(1-\kappa)e^{\beta_1y+\beta_0}}{\kappa e^{\beta_1y+\beta_0}}\\
&= \frac{1}{\kappa} e^{-\beta_1y-\beta_0}+ \frac{1-\kappa}{\kappa}.
\end{align}

From Equation \ref{eqn:exfam} applied to a mixture of normals,
\begin{eqnarray}
\label{app:norm}
Q(\eta,\beta) &=&\frac{1}{1+\int \sum_k^K w_kh(y)g(\eta)e^{\eta T(y)}(\frac{1}{\kappa} e^{-\beta_1y-\beta_0}+ \frac{1-\kappa}{\kappa})}\\
\nonumber &=& \frac{1}{1+e^{-\beta_0}\left(\sum_k^K \frac{w_k}{\kappa}\frac{g(\eta)}{g(\eta^*)}\right)+\frac{1-\kappa}{\kappa}}
\end{eqnarray}
We can invert Equation \ref{app:norm} to express $\beta_0$ as a function of  $Q$,$\eta$, $\beta$ and $\kappa$: 
\begin{eqnarray}
\label{app:norm1}
\beta_0(\eta,\beta_1,\kappa) &=& -\log\left(
                                 \frac{1-Q}{Q}\frac{(1+\frac{1-\kappa}{\kappa})}{                               \left(\sum_k^K \frac{w_k}{\kappa}\frac{g(\eta)}{g(\eta^*)}\right)} \right).
\end{eqnarray}
Finally using Equation \ref{miss_dist} we find that the missing data
are a mixture normals with weights
\begin{align}
\kappa^* &= \frac{\sum_k^K w_k\frac{g(\eta_k)}{g(\beta+\eta_k)}}{\sum_k^K
  w_k\frac{g(\eta_k)}{g(\beta+\eta_k)}+1-\kappa}\\
w_k^* &= \frac{w_k\frac{g(\eta_k)}{g(\eta_k^*)}}{\sum_k^K w_k\frac{g(\eta_k)}{g(\eta_k^*)}}.
\end{align}

\end{document}